\documentclass[12pt]{article}

\usepackage[centertags]{amsmath}
\usepackage{amsmath}
\usepackage{amsfonts}
\usepackage{amssymb}
\usepackage{amsthm}
\usepackage{newlfont}
\usepackage{amscd}
\usepackage{empheq}
\usepackage{epsfig,graphics,graphicx,amsfonts,psfrag}
\usepackage{verbatim}
\usepackage{eucal}

\newcommand{\tr}{\mbox{tr}}
\newcommand{\NN}{{\mathbb N}}

\newcommand{\RR}{{\mathbb R}}
\newcommand{\ZZ}{{\mathbb Z}}

\newcommand{\CC}{{\mathbb C}}
\newcommand{\beq}{\begin{equation}}
\newcommand{\eeq}{\end{equation}}
\newcommand{\ba}{\begin{array}}
\newcommand{\ea}{\end{array}}
\newcommand{\bea}{\begin{eqnarray}}
\newcommand{\eea}{\end{eqnarray}}
\newcommand{\eps}{{\epsilon}}

\usepackage{graphicx}
\usepackage[usenames,dvipsnames]{color}
\usepackage{transparent}
\usepackage{float}

\begin{document}

\begin{center}
{\bf Modulation instability, periodic anomalous wave recurrence, \\ and blow up in the Ablowitz - Ladik lattices}  

\vskip 10pt
{\it F. Coppini $^{1,2,3}$ and P. M. Santini $^{1,4}$}

\vskip 10pt

{\it 

$^1$ Dipartimento di Fisica, Universit\`a di Roma "La Sapienza", and 
Istituto Nazionale di Fisica Nucleare (INFN), Sezione di Roma, 
Piazz.le Aldo Moro 2, I-00185 Roma, Italy \\
$^2$ Department of Mathematics, Physics and Electrical Engineering, Northumbria University Newcastle,
Newcastle upon Tyne NE1 8ST, United Kingdom
}
\vskip 10pt

$^{3}$e-mail:  {\tt francesco.coppini@uniroma1.it,\\ francesco.coppini@roma1.infn.it}\\
$^{4}$e-mail:  {\tt paolomaria.santini@uniroma1.it, paolo.santini@roma1.infn.it}
\vskip 10pt

{\today}

\end{center}

\begin{abstract}

The Ablowitz-Ladik equations, hereafter called $AL_+$ and $AL_-$, are distinguished integrable discretizations of respectively the focusing and defocusing nonlinear Schr\"odinger (NLS) equations. In this paper we first study the modulation instability of the homogeneous background solutions of $AL_{\pm}$ in the periodic setting, showing in particular that the background solution of $AL_{-}$ is unstable under a monochromatic perturbation of any wave number if the  amplitude of the background is greater than $1$, unlike its continuous limit, the defocusing NLS. Then we use Darboux transformations to construct the exact periodic solutions of $AL_{\pm}$ describing such instabilities, in the case of one and two unstable modes, and we show that the solutions of $AL_-$ are always singular on curves of spacetime. At last, using matched asymptotic expansion techniques, we describe in terms of elementary functions how a generic periodic perturbation of the background solution i) evolves according to $AL_{+}$ into a recurrence of the above  exact solutions, in the case of one and two unstable modes, and ii) evolves according to $AL_{-}$ into a singularity in finite time if the  amplitude of the background is greater than $1$. The quantitative agreement between the analytic formulas of this paper and numerical experiments is perfect.

\end{abstract}

\section{Introduction}

The Ablowitz-Ladik (AL) equations \cite{AL1,AL2}:
\beq\label{AL}
\ba{l}
i\, \dot{u}_n+u_{n+1}+u_{n-1}-2u_n+\eta |u_{n}|^{2}\left(u_{n-1}+u_{n+1} \right) =0, \ \ \eta=\pm 1, \\
u_n=u(n,t)\in\CC, \ \ \dot{u}_n=\frac{du_n(t)}{dt}, \ \  n\in\ZZ, \ \ t\in\RR,
\ea
\end{equation}
are distinguished examples of integrable nonlinear differential-difference equations reducing, in the natural continuous limit 
\beq\label{limit}
u_n(t)=ih~v(\xi,\tau), \ \ h n =\xi, \ \ \tau =h^2 t, \ \ h\to 0, 
\eeq
to the celebrated integrable \cite{ZakharovShabat} nonlinear Schr\"odinger (NLS) equations
\beq\label{NLS}
\ba{l}
iv_{\tau}+v_{\xi\xi}+2\eta |v|^2v=0,  \ \eta=\pm 1, \\
v(\xi,\tau )\in\CC, \ \ v_{\tau}=\frac{\partial v}{\partial \tau}, \ \ v_{\xi\xi}=\frac{\partial^2 v}{\partial {\xi}^2}, \ \ \xi,\tau\in\RR,
\ea
\eeq
where $h$ is the lattice spacing. The two cases $\eta=\pm 1$ in \eqref{NLS} distinguish between the very different focusing ($\eta=1$) and defocusing ($\eta=-1$) NLS regimes. 

The AL equations \eqref{AL} characterize \cite{ItsKorepin} the quantum correlation function  of the XY-model of spins \cite{Lieb}. If $\eta=1$, it is relevant  in the study of anharmonic lattices \cite{Takeno}; it is gauge equivalent to an integrable discretization of the Heisenberg spin chain \cite{ishimori}, and appears in the description of a lossless nonlinear electric lattice ($\eta=1$) \cite{Marquie}. At last, if $\eta=1$, the AL hierarchy describes the integrable motions of a discrete curve on the sphere \cite{Doliwa}.

The well-known Lax pair of equations (\ref{AL}) reads \cite{AL1,AL2}
\beq\label{AL_Lax_1}
\ba{l}
\vec \psi_{n+1}(t,\lambda)=L_n(t,\lambda)\vec \psi_{n}(t,\lambda), \ \ \ \ \vec{\psi_{n}}_t(t,\lambda)=A_n(t,\lambda)\vec \psi_{n}(t,\lambda),\\
L_n(t,\lambda)=\begin{pmatrix}
\lambda & u_n(t) \\[2mm]
-\eta\overline{u}_n(t) & \frac{1}{\lambda}
\end{pmatrix}, \\
A_n(t,\lambda)=i\,\begin{pmatrix}
\lambda^2-1+\eta u_n \overline{u}_{n-1} & \lambda u_n-\frac{u_{n-1}}{\lambda} \\[2mm]
\eta\frac{\overline{u}_n}{\lambda}-\eta\lambda \overline{u}_{n-1}  & 1-\frac{1}{\lambda^2}-\eta {u}_n \overline{u}_{n-1}
\end{pmatrix},
\ea
\eeq
where $\bar f$ is the complex conjugate of $f$, and the matrices $L_n$ and $A_n$ of the Lax pair (\ref{AL_Lax_1}) possess the two symmetry
\beq\label{symmetries}
\ba{l}
L_n(\lambda)=P_\eta\, \overline{L_n\left(\frac{1}{\overline{\lambda}}\right)}P_\eta^\dagger=-\sigma_3\,L_n(-\lambda)\,\sigma_3, \\
A_n(\lambda)=P_\eta\, \overline{A_n\left(\frac{1}{\lambda^*}\right)}P_\eta^\dagger=\sigma_3\,A_n(-\lambda)\,\sigma_3,
\ea
\eeq
where \\
\beq\label{def_P}
\sigma_3=\begin{pmatrix}
  1 & 0 \\
  0 & -1
\end{pmatrix}, \ \ \ \ 
P_\eta=\begin{pmatrix}
0&-\eta\\1&0
\end{pmatrix}
\eeq
implying that, if $\vec\psi_n(\lambda,t)=({\psi_1}_n(\lambda,t),{\psi_2}_n(\lambda,t))^T$ is solution of (\ref{AL_Lax_1}), then
\beq\label{symmetries_psi}
\check{\vec\psi}_n(\lambda,t)=\left(
\ba{c}
-\eta\overline{{\psi_2}_n\left(\frac{1}{\overline{\lambda}},t)\right)}\\
\overline{{\psi_1}_n\left(\frac{1}{\overline{\lambda}},t)\right)}
\ea
\right), \ \ \hat{\vec\psi}_n(\lambda,t)=(-1)^n\left(
\ba{c}
{\psi_1}_n(-\lambda,t)\\
-{\psi_2}_n(-\lambda,t)
\ea
\right)
\eeq
are also a solution of (\ref{AL_Lax_1}).

The Inverse Scattering Transform \cite{AS,ZMNP} of the AL equations \eqref{AL} for localized initial data was developed in \cite{AL1,AL2} (see also \cite{APT}), and for non zero boundary conditions and $\eta=1$ in \cite{Prinari}. The Finite Gap method (FG) \cite{Novikov,Dubrovin,ItsMatveev,Krichever,ItsKotlj}  for periodic and quasi-periodic AL solutions was developed in \cite{Miller}.    

It is well-known that the homogeneous background solutions of the NLS equations \eqref{NLS}
\beq\label{background}
a \exp(2\,i\,\eta\,|a|^2\,\tau), \ \ a \ \mbox{complex constant parameter},
\eeq
is unstable under the perturbation of waves with sufficiently large wave length in the focusing case $\eta=1$ \cite{Talanov,BF,Zakharov,ZakharovOstro}, and always stable in the defocusing case $\eta=-1$, and the modulation instability (MI) of the focusing case is the main cause for the formation of anomalous (rogue) waves (AWs) \cite{HendersonPeregrine,Dysthe,Osborne,KharifPeli1,KharifPeli2,Onorato2,ZakharovOstro}. Since (\ref{background}) is also the exact homogeneous solution of the AL equations (replacing $\tau$ by $t$), it is natural to investigate their linear instability properties under monochromatic perturbations with respect to the AL dynamics, and study how such instability develops into the full nonlinear regime. 

We remark that, as in the NLS case, the AL equations have the elementary gauge symmetry (if $u$ is solution, also $\tilde u_n=u_n\exp(i\rho),~\rho\in\RR$ is solution); then $a$ could be chosen to be positive without loss of generality (but we shall not do it here). Unlike the NLS case, for which, if $v(\xi,\tau)$ is a solution, also $\tilde v(\xi,\tau)=b\, v(b \xi, b^2 \tau), ~b\in\RR$ is solution), the AL equations do not possess any obvious scaling symmetry.  It follows that $a$ in (\ref{background}) cannot be rescaled away as in the NLS case. Therefore one expects that, unlike the NLS case, the amplitude $a$ of the background (\ref{background}) play a crucial role in its stability properties under perturbation.

The Cauchy problem of the periodic AWs of the focusing NLS equation (\ref{NLS}) has been solved in \cite{GS1,GS2}, to leading order and in terms of elementary functions, for generic periodic initial perturbations of the unstable background:
\beq\label{eq:nls_cauchy1}
\ba{l}
v(\xi,0)=a\left(1+\varepsilon w(\xi)\right), \ 0<\varepsilon\ll 1, \ w(\xi+L)=w(\xi), \\
w(\xi)=\sum_{j=1}^{\infty}(c_j e^{i k_j \xi}+c_{-j}e^{-i k_j \xi}),\ \ k_j=\frac{2\pi}{L}j,
\ea
\eeq
in the case of a finite number $N$ of unstable modes, using a suitable adaptation of the finite-gap (FG) method. In the simplest case of a single unstable mode ($N=1$), the above finite gap solution provides the analytic and quantitative description of an ideal Fermi-Pasta-Ulam-Tsingou (FPUT) recurrence \cite{FPU} without thermalization, of periodic NLS AWs over the unstable background (\ref{background}),  described, to leading order, by the well-known Akhmediev breather (AB)
\beq\label{Akhmediev}
\ba{l}
{\cal A}(\xi,\tau;k,\tilde X,\tilde T,\rho)=\tilde a e^{2i{|\tilde a|}^2\tau+i\rho}\, \frac{\cosh\left[\tilde\sigma(k) (\tau-\tilde T)+2i\phi\right]+\sin\phi \cos[k(\xi-\tilde X)]}{\,\cosh[\tilde\sigma(k) (\tau-\tilde T)]-\sin\phi \cos[k(\xi-\tilde X)]}, \\
\tilde\sigma(k):=k\sqrt{4 |\tilde a|^2-k^2},
\ea
\eeq
solution of focusing NLS for the arbitrary real parameters $\tilde a,\rho,k,\tilde X,\tilde T$, but with different parameters at each appearance \cite{GS1}. See also \cite{GS3} for an alternative and effective approach to the study of the AW recurrence in the case of a single unstable mode, based on matched asymptotic expansions; see \cite{GS4} for a finite-gap model describing the numerical instabilities of the AB and \cite{GS5} for the analytic study of the linear, nonlinear, and orbital instabilities of the AB within the NLS dynamics; see \cite{GS6} for the analytic study of the phase resonances in the AW recurrence; see \cite{San} and \cite{CS2} for the analytic study of the FPUT AW recurrence in other NLS type models: respectively the PT-symmetric NLS equation \cite{AM1} and the massive Thirring model \cite{Thirring,Mikhailov}. The AB, describing the nonlinear instability of a single mode, and its $N$-mode generalization were first derived respectively in \cite{Akhmed0}  and \cite{ItsRybinSall}. The NLS recurrence of AWs in the periodic setting has been investigated in several numerical and real experiments, see, f.i., \cite{Yuen1,Yuen3,Kimmoun,Mussot,PieranFZMAGSCDR}, and qualitatively studied in the past via a 3-wave approximation of NLS \cite{Infeld,trillo3}. 

In addition, a perturbation theory describing analytically how the FPUT recurrence of AWs is modified by the presence of a perturbation of NLS has been recently introduced in \cite{CGS}, in the simplest case of a small linear loss or gain, giving a theoretical explanation of previous interesting real and numerical experiments \cite{Kimmoun,Soto}. This theory has been applied in \cite{CS1} to the complex Ginzburg-Landau \cite{Newell_Whitehead} and Lugiato-Lefever \cite{LL} models, treated as perturbations of NLS (see also \cite{CGS2}).

These results suggest two interesting problems.\\
i) the construction of the analytic and quantitative description of the dynamics of periodic AWs of the AL lattices;\\
ii) the understanding of the effect of a perturbation of the AL lattices on such a dynamics.

The solution of the first problem is the main goal of this paper; the solution of second problem is contained in the paper \cite{CS3}.

We remark that the terminology ``focusing'' and ``defocusing'' NLS equations should not be exported to their AL discretizations, since the background solution of the AL equation reducing to the defocusing NLS is unstable under any monochromatic perturbation if $|a|>1$, and such an instability leads generically to blow up in space-time. Therefore we prefer to call hereafter the AL equations \eqref{AL} with $\eta=\pm 1$ as $AL_{\pm}$ equations, instead of using the terminology ``focusing and defocusing AL equations'' often present in the literature.      

The paper is organized as follows. In \S 2 we investigate the linear stability properties of the background \eqref{background}, extending results already present in the literature \cite{Akhm_AL,Otha}, and showing that, unlike the NLS case, the background of $AL_-$ is unstable under any monochromatic perturbation if $|a|>1$. In \S 3 we present the exact solutions of $AL_{\pm}$ describing the instability of one and two unstable modes, and we show that: i) the solutions of $AL_{+}$ are always regular, but with an amplitude, relative to the background, growing as $|a|^2$ and as $|a|^4$ respectively in the case of one and two unstable modes; ii) the solutions of $AL_{-}$ develop singularities in closed curves of spacetime, and when these curves intersect a line $x=n_0\in\ZZ$ (the generic case), the solution blows up at finite time in the site $n=n_0$. In \S 4 we use the matched asymptotic expansion approach developed in \cite{GS3} to solve to leading order the periodic Cauchy problem of the AWs, i) describing in terms of elementary functions the associated AW recurrence of one and two unstable modes (in this second case for non generic initial perturbations) in the  $AL_{+}$ model; ii) showing analytically how a smooth perturbation blows up at finite time in the $AL_{-}$ model. The Appendix is dedicated to the construction of the exact solutions studied in \S 3 using the Darboux transformations (DTs) of $AL_{\pm}$ \cite{Geng}.  

 To the best of our knowledge, known results concerning AWs of the $AL_{\pm}$ models prior to our work are the following. 
The exact solution of $AL_+$ over the background, corresponding to a spectral parameter in general position, containing as limiting cases the discrete analogues of the Akhmediev breather \eqref{Narita1}, of the Kuznetsov-Ma \cite{Kuznetsov,Ma} and Peregrine \cite{Peregrine} solutions were first constructed in \cite{Narita} using the Hirota method \cite{Hirota0}. The amplitude growth of \eqref{Narita1} for large $|a|$ was investigated in \cite{Akhm_AL}, together with the linear instability properties of the background solution of $AL_+$. Numerical experiments for the Cauchy problem of AWs for $AL_+$ are reported in \cite{Schober}. The linear instability properties of the background solution of the $AL_-$ model were investigated in \cite{Otha}, where Peregrine type solutions of any order of the $AL_{\pm}$ models were constructed using the Hirota method, observing that they are singular in the $AL_-$ case.

\section{Linear stability properties of the background in the AL case}

To study the (linear) stability properties of the background solutions (\ref{background}) in the $AL_{\pm}$ dynamics, we seek solutions of (\ref{AL}) in the form
\beq\label{monochr_pert}
u_n(t)=a e^{2\,i\,\eta\,|a|^2\,t}\left(1+\eps \xi_n(t) +O(\eps^2)\right), \ \ 0<\eps\ll 1,
\eeq
obtaining, at $O(\eps)$, the linearized $AL_{\pm}$ equations for $\xi_n$:
\beq\label{linearized}
i\dot{\xi}_n+(1+\eta |a|^2)(\xi_{n+1} +\xi_{n-1})-2\xi_n+2\eta |a|^2\overline{\xi_n}=0, \ \ \eta=\pm 1.
\eeq
If the perturbation is a monochromatic wave:
\beq\label{perturbation1}
\xi_n(t)=\gamma_+(t)e^{i\kappa n}+\gamma_{-}(t)e^{-i\kappa n},
\eeq
equation (\ref{linearized}) reduces to the system of ODEs
\beq\label{equ_sum_diff}
\ba{l}
i\dot D+2[(1+\eta |a|^2)\cos\kappa-(1-\eta |a|^2)]S=0, \\
i\dot S+2(1+\eta |a|^2)(\cos\kappa-1)D=0, \\
S:=\gamma_+ +\overline{\gamma_{-}}, \ \ \ D:=\gamma_+ -\overline{\gamma_{-}}.
\ea
\eeq
whose solution reads:
\beq\label{sol_S_D}
S(t)=\nu e^{\sigma t}+\mu e^{-\sigma t}, \ \ D(t)=\frac{i\sigma(\nu e^{\sigma t}-\mu e^{-\sigma t})}{2(1+\eta |a|^2)(1-\cos\kappa)},
\eeq
where
\beq\label{def_sigma}
\sigma(\kappa)=2\sqrt{(1+\eta |a|^2)(1-\cos\kappa)\left[(1+\eta |a|^2)\cos\kappa-(1-\eta |a|^2)\right]},
\eeq
and $\nu,\mu$ are two arbitrary complex parameters.
From now on we fix the following constraint on the wave number
\beq
0<\kappa<\pi ,
\eeq
since the negative values are covered by the second exponential in (\ref{monochr_pert}), and the growth rate (\ref{def_sigma}) depends on $\kappa$ through $\cos\kappa$.  

\begin{center}
\begin{figure}[H]
	  \includegraphics[width=6cm]{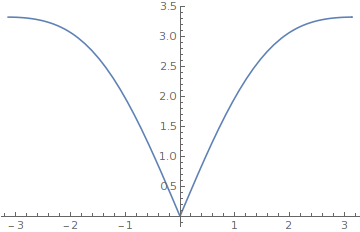}
	  \includegraphics[width=6cm]{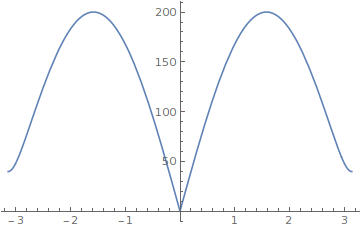}
	\caption{For $\eta=-1$ and $|a|>1$ the perturbation is unstable with exponential growth. The graph of the  growth rate $\sigma(\kappa)$ in the basic period $\kappa\in(-\pi,\pi)$. In the left picture: $a=1.3<\sqrt{2}$; in the right one: $|a|=10>\sqrt{2}$.}
	\label{fig:aa1}
\end{figure}
\end{center}

The growth rate (\ref{def_sigma}) implies the following stability features of the background solutions (\ref{background}) of the $AL_{\pm}$. \\
{\bf The case $\eta=-1$}. Equation $AL_{-}$ reduces to the defocusing NLS in the continuous limit, for which the background (\ref{background}) is stable under a perturbation of any wave number. The stability properties of the $AL_-$ background are much richer \cite{Otha}; indeed we distinguish three sub-cases:
\begin{itemize}
\item $|a|>1$ \ $\Rightarrow$ \ $\sigma>0$ \ $\Rightarrow$ \ exponential growth and linear instability $\forall \ \kappa$ (see Figure~\ref{fig:aa1}); 
\item $|a|=1$ \ $\Rightarrow$ \ $\sigma=0$, \ $S(t)=S_0, \ D(t)=-4iS_0t+D_0$ \ $\Rightarrow$ \ instability if $S_0\ne 0$ with linear growth, stability otherwise, \ $\forall \ \kappa$; 
\item $0<|a|<1$ \ $\Rightarrow$ \ $\sigma\in i\RR$ \ $\Rightarrow$ \ oscillations and neutral stability $\forall \ \kappa$. 
\end{itemize}

Since $\sigma$ depends on $\kappa$ through $\cos\kappa$, these stability properties are $2\pi$-periodically extended to the whole real $\kappa$ axis, with basic period $(-\pi,\pi)$. In the unstable case $|a|>1$, there are two subcases:\\
i) $1<|a|<\sqrt{2}$, then $\sigma(k)$ has its absolute minimum at $\kappa=0$ with $\sigma(0)=0$, and absolute maxima at $\kappa=\pm\pi$, with $\sigma(\pm\pi)=4\sqrt{|a|^2 -1}$. \\
ii) $|a|>\sqrt{2}$, then $\sigma(k)$ has its absolute minimum at $\kappa=0$ with $\sigma(0)=0$, relative minima at $\kappa=\pm\pi$, with $\sigma(\pm\pi)=4\sqrt{|a|^2 -1}$, and absolute maxima at $\kappa_{\pm}=\pm\arccos\left(\frac{1}{1-|a|^2}\right)$, with $\sigma(\kappa_{\pm})=2 |a|^2$ (see Figure~\ref{fig:aa1}).
\textbf{The case $\eta=1$}. The $AL_+$ equation reduces to the focusing NLS (\ref{NLS}) in the continuous limit, for which the background (\ref{background}) is unstable for monochromatic perturbations of wave number $k$ such that $|k|<2 |a|$, and the parameter $a$ can be rescaled away due to the scaling and trivial gauge symmetries of NLS. Also in this case the stability properties of the AL background are richer than those of the NLS background, since now the amplitude $|a|$ cannot be rescaled away, and is involved in the following nontrivial way.

\begin{figure}[H]
\begin{center}
	\includegraphics[width=6.5cm]{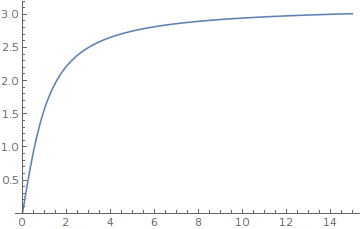} \ \includegraphics[width=6.5cm]{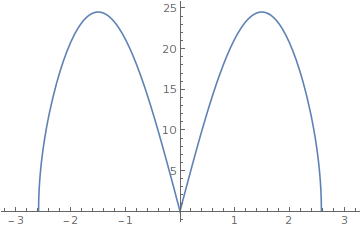}
	\caption{\label{fig:aa2} Left: the critical wave number $\kappa_a$ as function of the amplitude $|a|$, with $\kappa_0=0$ and $\kappa_{\infty}=\pi$. Right: the growth rate, $2\pi$-periodic in $\kappa$, has support for $|\kappa|<\kappa_a$ inside the basic periodicity interval $(-\pi,\pi)$. Here $|a|=3.5$.}
\end{center}
\end{figure}

Define $\kappa_a$ as \cite{Akhm_AL}
\beq
\kappa_a:=\arccos\left(\frac{1-|a|^2}{1+|a|^2} \right)>0, \ \ 0<\kappa_a <\pi ;
\eeq
(see Figure~\ref{fig:aa2}); then
\begin{itemize}
\item if $|\kappa |<\kappa_a$ \ $\left(\cos\kappa >\frac{1-|a|^2}{1+|a|^2}\right)$, \ $\sigma>0$ \ $\Rightarrow$ \ instability with exponential growth. The growth rate has maxima at $\pm\kappa_M$, with $\kappa_M=\arccos\left(\frac{1}{|a|^2+1} \right)$ and $\sigma(\pm\kappa_M)=2|a|^2$. The instability curve is similar to that of focusing NLS, except for the $2\pi$ periodicity (see Figures~\ref{fig:aa2}).  
\item if $|\kappa |=\kappa_a$ \ ($\cos\kappa =\frac{1-|a|^2}{1+|a|^2}$) \ $\Rightarrow$ \ $\sigma=0$, \ $D=D_0,~S=-4i|a|^2D_0 t+S_0$ \ $\Rightarrow$ \  instability with linear growth if $D_0\ne 0$, otherwise stability;
\item if $|\kappa |>\kappa_a$ \ $\left(\cos\kappa <\frac{1-|a|^2}{1+|a|^2}\right)$ \ $\Rightarrow$ \ $\sigma\in i\RR$ \ $\Rightarrow$ \ linear stability and small oscillations.
\end{itemize}
As before, these stability properties are $2\pi$-periodically extended to the whole real $\kappa$ axis.

Summarizing the results of this section, we have the following ``instability regions'' of the background (\ref{background}):
\beq\label{unstable_cases}
\ba{ll}
|\kappa|<\kappa_a:=\arccos\left(\frac{1-|a|^2}{1+|a|^2}\right), \ \forall |a|>0, & \mbox{if } \eta=1,\\

|a|>1,\ \forall \kappa,  &  \mbox{if } \eta=-1,
\ea
\eeq
where $\kappa_a$ is the smallest positive branch of $\arccos$. They were found in \cite{Akhm_AL} for $\eta=1$, and in \cite{Otha} for $\eta=-1$. 

As we shall see in the following, in all the cases discussed above in which the background (\ref{background}) is unstable, we find it convenient to introduce the parameter $\phi$ defined by  
\beq\label{def_phi}
\cos  \phi= \sqrt{1+\frac{\eta}{|a|^2} }\,\sin \left(\frac{\kappa}{2}\right), \ \ \ \ 0<\phi<\pi/2.
\eeq
We remark that $\phi$ is real in both unstable cases (\ref{unstable_cases}) (it is therefore an angle) and, in terms of it, the growth rate $\sigma$ \eqref{def_sigma} takes the same simple form
\beq
\sigma=2 |a|^2 \sin( 2\phi )>0, 
\eeq
as in the NLS case \cite{GS1,GS2,GS3}.

It is easy to verify, from (\ref{equ_sum_diff}),(\ref{sol_S_D}), and (\ref{def_phi}), that the perturbation $\xi_n$ in (\ref{perturbation1}) reads
\beq\label{sol_lin}
\xi_n(t)=\frac{1}{\sin(2\eta\phi)}\Big[|\alpha|e^{\sigma t+i\eta\phi}\cos\left[\kappa(x-X^+) \right]+|\beta|e^{-\sigma t-i\eta\phi}\cos\left[\kappa(x-X^-) \right] \Big],
\eeq
where
\beq\label{def_X+_X-}
\ba{l}
X^+=\frac{\arg\alpha+\pi/2}{\kappa}, \ \ X^-=\frac{-\arg\beta+\pi/2}{\kappa},
\ea
\eeq
and the arbitrary complex parameters $\alpha,\beta$ are expressed in terms of $\nu,\mu$ as follows:
$\alpha=-2i\sin(\eta\phi)~\overline{\nu}, \ \ \beta=2i\sin(\eta\phi)~\mu$.

\section{Exact periodic AW solutions of the AL equations}

Since the background solution (\ref{background}) is linearly unstable under monochromatic perturbations in the cases (\ref{unstable_cases}), it is important to describe how the corresponding exponential growth evolves into the nonlinear stage of MI described by the full nonlinear model. In this section we present the exact periodic solutions of $AL_{\pm}$ describing the nonlinear instability of a single nonlinear mode and of two interacting nonlinear modes. The construction of these solutions using the DTs of the AL equations \cite{Geng} is presented in the Appendix.

\subsection{The case of a single unstable mode}

The instability of a single nonlinear mode $K_1$ of $AL_{\pm}$ is described by the solution:
\beq\label{Narita1}
{\cal N}_1(n,t;K_1,X_1,T_1,\rho,\eta)=ae^{2i\eta |a|^2t+i\rho}\, \frac{\cosh\left[\sigma(K_1) (t-T_1)+2i\eta\phi\right]+\eta G_1 \cos[K_1(n-X_1)]}{\,\cosh[\sigma(K_1) (t-T_1)]-\eta G_1 \cos[K_1(n-X_1)]}, 
\eeq
where
\beq\label{def_G1}
G_1=\frac{\sin \theta_1}{\cos\left( \frac{K_1}{2}\right)},
\eeq
$K_1$ is the wave number and $\sigma(K_1)$, defined in (\ref{def_sigma}), is the growth rate of the linearized theory in the unstable cases (\ref{unstable_cases}), the angle $\theta_1$ is defined as in (\ref{def_phi})
\beq\label{def_theta}
\cos  \theta_1 = \sqrt{1+\frac{\eta}{|a|^2} }\,\sin \left(\frac{K_1}{2}\right),
\eeq
and $X_1$, $T_1$ and $\rho$ are arbitrary real parameters. Since $\theta_1$ is defined in terms of ($K_1,a,\eta$), the growth rate $\sigma(K_1)$ and the parameter $G_1$ can be expressed in terms of $(K_1,a,\eta)$ or in terms of $(\theta_1,a,\eta$) in the following way:
\begin{equation}\label{def_Sigma1}
\ba{l}
\sigma(K_1)=2\sqrt{(1+\eta |a|^2)(1-\cos K_1)\left[(1+\eta |a|^2)\cos K_1-(1-\eta |a|^2)\right]}\\
=2 |a|^2 \sin( 2\theta_1 ), \\
G_1=\frac{\sin \theta_1}{\cos\left( \frac{K_1}{2}\right)}=\sqrt{1-\frac{\eta}{|a|^2}\frac{1-\cos K_1}{1+\cos K_1}}
=\frac{\sqrt{|a|^2+\eta}\sin\theta_1}{\sqrt{|a|^2\sin^2\theta_1 +\eta}}.
\ea
\end{equation}

If $\eta=1$, (\ref{Narita1}) is the Narita solution \cite{Narita} of $AL_+$, discrete analogue of the AB solution \eqref{Akhmediev} of focusing NLS, reducing to it through the scaling
\beq\label{limit_Nar_Akhm}
\ba{l}
a\sim h ~\tilde a, \ K_1\sim h ~k \ \Rightarrow \ \sigma(K_1)\sim h^2 \tilde\sigma(k), \ G_1\sim \sin\phi, \ \ h\ll 1, \\
\xi\sim h n, \ \tau\sim h^2 t, \ \tilde X\sim h X, \ \tilde T\sim h^2 T, \ \ h\ll 1,
\ea
\eeq
where $h$ is the lattice spacing. If $\eta=-1$, (\ref{Narita1}) is, to the best of our knowledge, the novel solution describing the MI present also in the $AL_{-}$ model.

\begin{figure*}[h!!] 
	\includegraphics[height=6 cm,width= 8 cm]{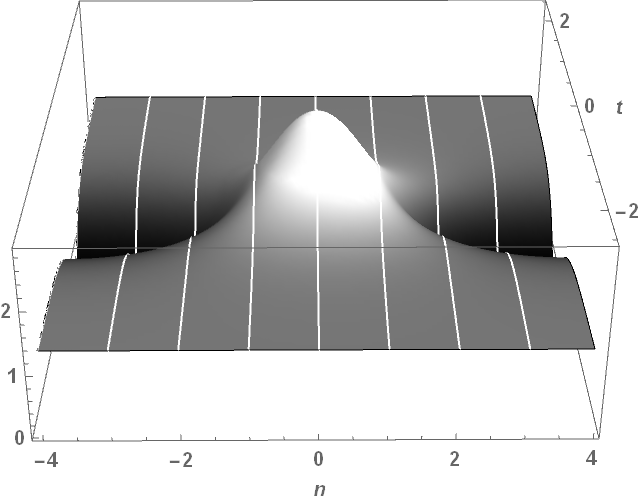} \caption{\label{Narita} 3D plot of $|{\cal N}_1|$  for $\eta=1$, $a=1.3$, $K_1=2\pi/8$, $X_1=T_1=0$.}
\end{figure*}

The solution (\ref{Narita1}) oscillates in $n$ and is exponentially localized over the background in $t$ in the following way
\beq\label{phase_shift}
{\cal N}_1(n,t;K_1,X_1,T_1,\rho,\eta)\rightarrow a e^{2\,i\,\eta |a|^2t\pm 2i\eta\phi+i\rho}, \;\;\;\; \mbox{as} \;\;\; t\rightarrow\pm\infty .
\eeq
To study its behavior, we first replace $n\in\ZZ$ by $x\in\RR$ in \eqref{Narita}; it is legitimate, observing that function ${\cal N}_1(x,t;K_1,X_1,T_1,\rho,\eta)$ solves $AL_{\pm}$ with $n$ replaced by $x$:
\beq\label{AL1}
  \ba{l}
  iu_t+u^++u^--2u+\eta |u|^2(u^++u^-)=0, \ \ \eta=\pm 1, \\
  u=u(x,t)\in\CC, \ \ u^{\pm}=u(x\pm 1,t), \ \ u_t=\frac{\partial u}{\partial t}, \ \  x,t\in\RR.
  \ea
  \eeq

If $\eta=1$, equation (\ref{def_theta}) implies that $\sin\theta_1<\cos(K_1/2)$ and equation (\ref{def_G1}) that $G_1<1$. It follows that the denominator of ${\cal N}_1$ is always positive. Therefore the solution ${\cal N}_1$ is always regular in the $(x,t)$ plane for all values of its arbitrary parameters, like in the NLS case. But there is an important difference, since now 
the maximum of the absolute value of the solution (\ref{Narita1}), reached at the point $(x,t)=(X_1,T_1)$, reads \cite{Akhm_AL}:
\begin{equation}
\ba{l}
Max:=\max\limits_{(x,t)\in\RR^2}|{\cal N}_1(x,t;K_1,X_1,T_1,\rho,1))|=|{\cal N}_1(X_1,T_1;K_1,X_1,T_1,\rho,1)|\\
=|a|\left|\frac{\cos(2\theta_1)+G_1}{1-G_1} \right|
=|a|\left[2\left(1+|a|^2\right)\cos\left(\frac{K_1}{2}\right)\left(\sin \theta_1 +\cos\left(\frac{K_1}{2}\right)\right)-1\right],
\ea
\end{equation}
implying that the relative maximum (the ratio of the maximum of the amplitude of (\ref{Narita1}) to the background amplitude $|a|$) grows as $O(|a|^2)$ for $|a|\gg 1$:
\beq
\frac{M}{|a|}=4 |a|^2 \cos^2\left(\frac{\kappa}{2}\right)\left[1+O(|a|^{-2})\right], \ \ a\gg 1,
\eeq
unlike the NLS case, for which $M/|a|=1+2\sin\phi$ does not depend on $a$. We remark that, if $X_1\notin\ZZ$, the maximum of $|{\cal N}_1|$ is not reached in a lattice point (see Figure \ref{Narita}).

\begin{figure*}[h]
  \begin{center}
    \includegraphics[height=6.5 cm,width= 13 cm]{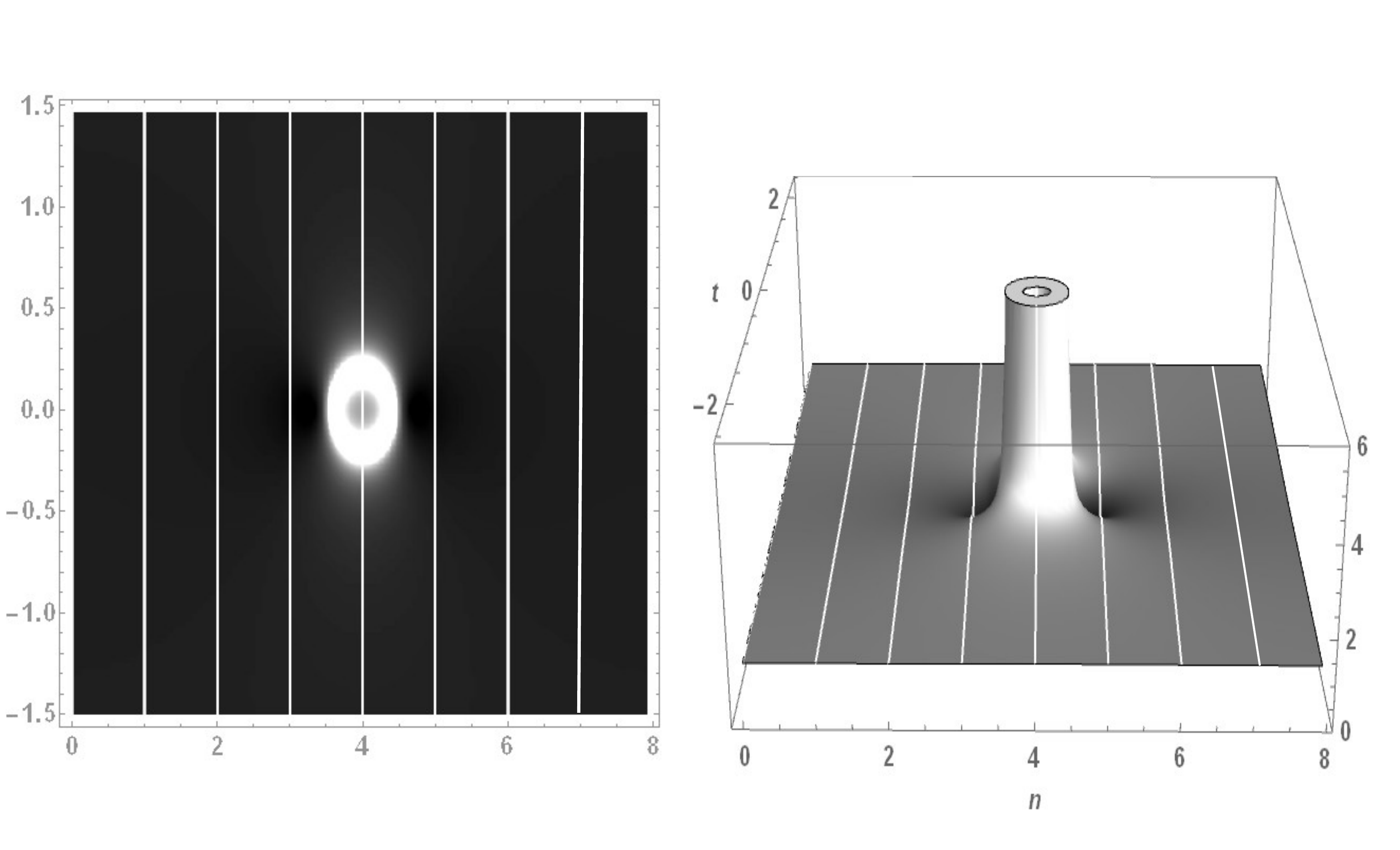} \caption{\label{Narita_defoc} Density and 3D plots of the singular solution $|{\cal N}_1|$ of $AL_{-}$, where $a=1.3$, $K_1=2\pi/8$, $T_1=0$, $X_1=4$.}
    \end{center}
\end{figure*}

If $\eta=-1$, equation (\ref{def_theta}) implies that $\sin\theta>\cos(K_1/2)$ and equation (\ref{def_G1}) that $G_1>1$. Therefore the solution \eqref{Narita1} is singular on the closed curve $\cal C$ of the $(x,t)$ plane defined by the equation 
\beq\label{singularity}
\cosh[\sigma (t-T_1)]=G_1\cos[\kappa(x-X_1)];
\eeq
this curve is centered at $(X_1,T_1)$ and $x$-periodic with period $2\pi/K_1$ (see Figures~\ref{Narita_defoc}). If
\beq
|x-X_1|<\frac{\arccos(1/G_1)}{K_1},
\eeq
the solution ${\cal N}_1$ blows up at the two points $(x,t^{\pm}(x))$:
\beq\label{sing_points}
t^{\pm}(x)=T_1\pm\frac{\cosh^{-1}\left(G\cos[\kappa(x-X_1)]\right)}{\sigma};
\eeq
where $\arccos$ is here the smallest positive branch of the inverse of $\cos$, and $\cosh^{-1}$ is the positive branch of the inverse of $\cosh$ (see Figures~\ref{Narita_defoc}).

We remark that the extension $l$ of the singular curve (\ref{singularity}) in the $x$ direction is less than $1$, since:
\beq
l=\frac{2}{K_1}\arccos\left(\frac{1}{G_1}\right)<1 \ \ \Leftrightarrow \ \ \frac{1}{\sin\theta_1}>1 .
\eeq
Consequently, if $n_X$ is the integer closest to $X_1$ and
\beq
|n_X -X_1|>\frac{\arccos\left(1/G_1\right)}{K_1},
\eeq
then the appearance of the AW is not singular on the lattice, since the singular curve is located in the region between two subsequent sites. But this situation is not generic.

We remark that the solution (\ref{Narita1}) for $\eta=-1$ does not have a continuous limit to defocusing NLS, since the prescription $a\sim h \tilde a$, $h\ll 1$ in \eqref{limit_Nar_Akhm} is not consistent with the instability condition $|a|>1$.

\subsection{The case of two unstable modes}

The instability of two nonlinearly interacting modes $K_1$ and $K_2$ is described by the novel (to the best of our knowledge) two-mode solution of $AL_{\pm}$:
\begin{equation}\label{Narita2} 
{\cal N}_2(n,t;K_1,K_2,X_1,X_2,T_1,T_2,\rho,\eta)=a\,e^{2i\eta |a|^2t+i\,\rho}\,\frac{N(n,t)}{D(n,t)},
\end{equation}\\[2mm]
where
\begin{equation} 
\ba{l}
N(n,t)=\cosh[\Sigma_1(t-T_1)+\Sigma_2(t-T_2)+2\,i\eta(\theta_1+\theta_2)]+\\[2mm]
\left(a_{12}(K_1,K_2)\right)^2\cosh[\Sigma_1(t-T_1)-\Sigma_2(t-T_2)+2\,i\eta(\theta_1-\theta_2)]\\[2mm]
+2\,a_{12}(K_1,K_2)\,\bigg\{G_2\cosh[\Sigma_1\,(t-T_1)+2i\eta\theta_1]\,\cos[K_2(n-X_2)]\\
+G_1\cosh[\Sigma_2\,(t-T_2)+2i\eta\theta_2]\,\cos[K_1(n-X_1)]
\bigg\}\\[2mm]
+b^-_{12}(K_1,K_2)\cos[K_1(n-X_1)+K_2(n-X_2)]\\
+b^+_{12}(K_1,K_2)\cos[K_1(n-X_1)-K_2(n-X_2)],
\ea
\eeq
\beq
\ba{l}
D(n,t)=\cosh[\Sigma_1(t-T_1)+\Sigma_2(t-T_2)]+\left(a_{12}(K_1,K_2)\right)^2\cosh[\Sigma_1(t-T_1)-\Sigma_2(t-T_2)]\\
-2\,a_{12}(K_1,K_2)\,\bigg\{G_2\cosh[\Sigma_1\,(t-T_1)]\,\cos[K_2(n-X_2)]\\
+G_1\cosh[\Sigma_2\,(t-T_2)]\,\cos[K_1(n-X_1)]
\bigg\}\\
+b^-_{12}(K_1,K_2)\cos[K_1(n-X_1)+K_2(n-X_2)]\\
+b^+_{12}(K_1,K_2)\cos[K_1(n-X_1)-K_2(n-X_2)],
\ea
\end{equation}
and where
\beq
\ba{l}
\cos \theta_j=\sqrt{1+\frac{\eta}{|a|^2} }\sin\left( \frac{K_j}{2}\right), \ \ j=1,2, \\[2mm] 
G_j=\frac{\sin(\theta_j)}{\cos\left(\frac{K_j}{2}\right)}, \ \ j=1,2, \\[2mm]
\Sigma_j=\sigma(K_j)=2 |a|^2 \sin( 2\theta_j ), \ \ \ j=1,2,
\ea
\eeq
\beq
\ba{l}
a_{12}=\frac{\sin(\theta_1+\theta_2)}{\sin(\theta_1-\theta_2)}, \\
b^{\pm}_{12}=\frac{\sin(\theta_1)\sin(\theta_2)}{\sin^2(\theta_1-\theta_2)}\left(\sqrt{\dfrac{\cos(\frac{K_2}{2})}{\cos(\frac{K_1}{2})}}\cos(\theta_1)\pm\sqrt{\dfrac{\cos(\frac{K_1}{2})}{\cos(\frac{K_2}{2})}}\cos(\theta_2)\right).
\ea
\eeq
Also this solution oscillates in $n$ and is exponentially localized in time over the background:\\[2mm]
\begin{equation*} 
{\cal N}_2(n,t;K_1,K_2,X_1,X_2,T_1,T_2,\rho,\eta)\rightarrow a\,e^{2i\eta |a|^2t+i[\rho\pm 2\eta(\theta_1+\theta_2 )]}, \;\;\;\; as \;\; t\rightarrow\pm\infty .
\end{equation*}\\[2mm]
In the rest of the paper we shall limit our considerations to the case $K_2=2 K_1$; then the solution is periodic with period $2\pi/K_1$.

As in the case of a single mode, in the natural continuous limit (see \eqref{limit_Nar_Akhm}) the solution for $\eta=1$ reduces to the two breather solution of Akhmediev type \cite{GS3}, while it does not have a continuous limit in the case $\eta=-1$.

\begin{figure}[H]
  \begin{center}
  \includegraphics[height=4.5 cm,width= 7.5 cm]{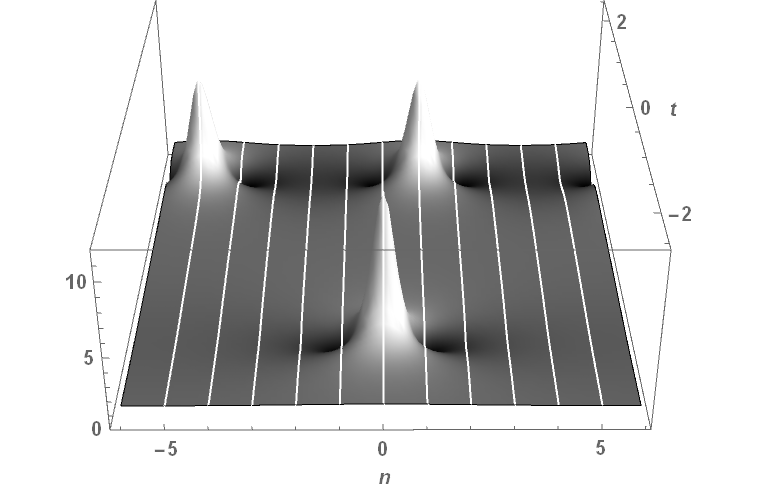}\includegraphics[height=4.5 cm,width= 7.5 cm]{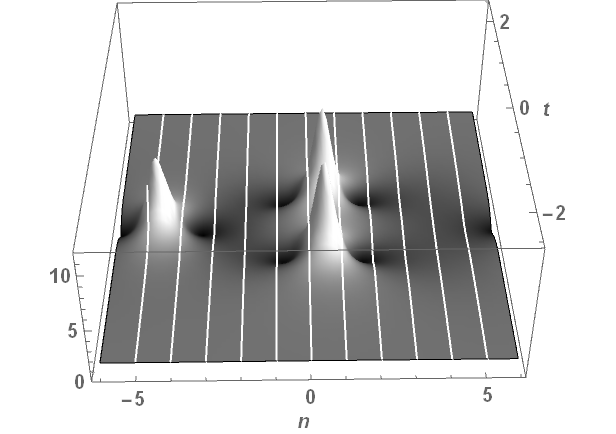}\\
\includegraphics[height=7 cm,width= 9 cm]{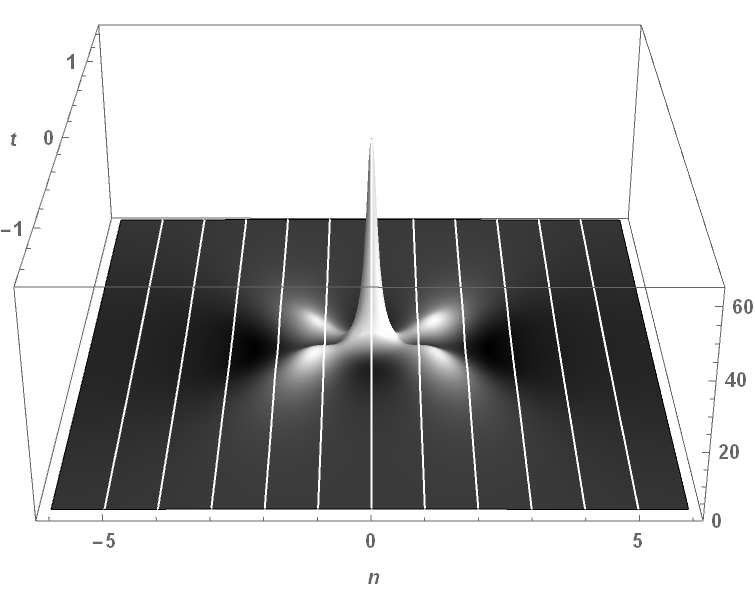}
\end{center}
\caption{\label{fig_reg} 3D plots of $|{\cal N}_2|$ for $\eta=1$, $a=1.3$, $K_1=2\pi/12$, and $K_2=2K_1$. Top left: $T_1=-1$, $T_2=1.5$, $X_1=0$ and $X_2=1$. $|T_2-T_1|\ge 1$, and the solution appears as two separate Narita solutions of wave numbers $K_1$ and $K_2$. Top right: $T_1=T_2=0$, $X_1=0$ and $X_2=1$. Since $T_1=T_2$, the two modes appear together and strongly interact. Bottom: the phase locking choice of the parameters: $T_1=T_2=0$, $X_1=0$ and $X_2=X_1+\frac{2\pi}{4 \,K_1}=3$.}
\end{figure}

As in the case of a single mode, it would be possible to show the following. \\
i) If $\eta=1$, the solution (\ref{Narita2}) is always regular. If $|T_1-T_2|>O(1)$ the two nonlinear modes are separated into two weakly interacting Narita solutions with wave numbers $K_1$ and $K_2$. If $|T_1-T_2|\ll 1$ the two nonlinear modes appear almost at the same time interacting nonlinearly. If $T_1=T_2$, $K_2=2 K_1$, and $X_2 =X_1 +\frac{2\pi}{4 \,K_1}$ the two modes are amplitude-locked and phase-locked in a characteristic configuration similar to the one of NLS (see Figures \ref{fig_reg})).

The maximum height of $|{\cal N}_2|$ can be calculated in terms of elementary functions in two cases: when $|T_1-T_2|\gg 1$ and the solution describes two separated Narita solutions (\ref{Narita1}), and when they are amplitude- and phase-locked:
\beq
T_1=T_2, \ \ \ \ \ X_2=X_1+\frac{2\pi}{4 \,K_1} .
\eeq
In the second case, the maximum height reached by the solution is given by:
\begin{equation} 
\ba{l}
\max \;|{\cal N}_2|=|{\cal N}_2(X_1,T_1; K_1,K_2,X_1,X_1\pm\frac{2\pi}{4\,K_1},T_1,T_1,\rho,1)|\\[2mm]
=|a|\left[2(1+|a|^2)\cos\left(\frac{K_1}{2}\right)\left(\cos(\theta_1)+\cos\left(\frac{K_1}{2}\right)\right)-1 \right] \\
\times \bigg[2(1+|a|^2)\cos\left(\frac{K_2}{2}\right)\left(\cos(\theta_2)+\cos\left(\frac{K_2}{2}\right)\right)-1 \bigg],
\ea
\end{equation}\\[3mm]
implying that the maximum amplitude of $|{\cal N}_2|$, relative to the background, grows as $|a|^4$ for $|a|$ large:  $\max |{\cal N}_2|/|a|=O\left(|a|^4\right), \ \ |a|\gg 1$ (see the last of Figures \ref{fig_reg}). 

\begin{figure*}[h!!] 
  \includegraphics[height=4.6 cm,width= 7.6 cm]{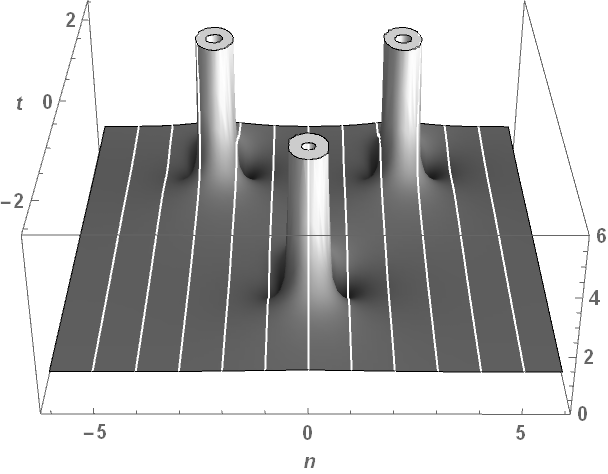}\includegraphics[height=4.6 cm,width= 7.6 cm]{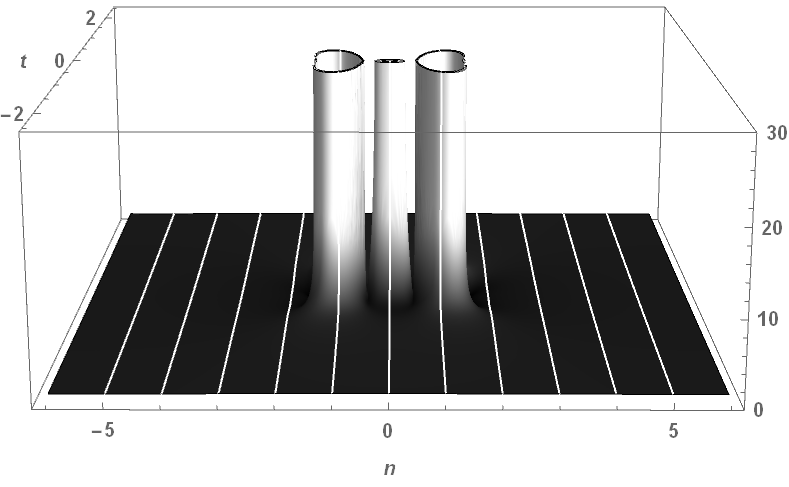}

  \caption{\label{singg} 3D plots of  $|{\cal N}_2|$ for $\eta=-1$, $a=1.3$, $K_1=2\pi/12$, $K_2=2K_1$. Left: $T_1=0$, $T_2=1$, $X_1=0$ and $X_2=3$. Since $|T_1-T_2|=1$, the solution describes two separate weakly interacting singular solutions of the type ${\cal N}_1$, with modes $K_1$ and $K_2=2 K_1$. Right: $T_1=T_2=0$, $X_1=0$ and $X_2=X_1+\frac{2\pi}{4\,K_1}=3$ (the phase-locking choice of the parameters)}.
\end{figure*}

\newpage

\noindent
ii) If $\eta=-1$, ${\cal N}_2$ develops always singularities at finite time on three closed curves of the $(x,t)$ plane; one curve for the mode $K_1$ and two curves for $K_2$ (see Figures \ref{singg}).

\section{AW recurrence and blow up at finite time of periodic AWs of $AL_{\pm}$}

In this section we study the periodic Cauchy problem with period $M\in\NN^+$ for $AL_{\pm}$ 
\beq
\ba{l}
i\, \dot{u}_n-2u_n+\left(1+\eta |u_{n}|^{2}\right)\left(u_{n-1}+u_{n+1} \right) =0, \\[2mm]
u_{n+M}(t)=u_n(t), \ \ \forall n\in\ZZ, \ \forall t\ge 0,
\ea
\eeq
in which the initial condition is a generic periodic perturbation of the background (\ref{background}) (what we call the ``periodic AW Cauchy problem''):
\beq\label{Cauchy}
\ba{l}
u_n(0)=a\left(1+\eps \left(\sum\limits_{j=1}^{p}\left(c_j e^{i \kappa_j n}+c_{-j}e^{-i \kappa_j n}\right)+c_0\right)\right), \ \ \ 0<\eps\ll 1, 
\ea
\eeq
where
\beq
\kappa_j=\frac{2\pi}{M}j, \ \ \ 1\le j \le p,
\eeq
and
\beq
p=\left\{
\ba{ll}
\frac{M}{2}, & \mbox{if $M$ is even}, \\
\ & \ \\
\frac{M-1}{2}, & \mbox{if $M$ is odd}.
\ea\right.
\eeq
As we shall see in the following, it is convenient to define the parameters
\beq\label{def_sigmaj_X+_X-}
\ba{l}
\sigma_j=2 a^2 \sin(2\phi_j), \\[2mm]
X^+_j=\frac{\arg(\alpha_j)+\pi/2}{\kappa_j}, \ \ \ X^-_j=\frac{-\arg(\beta_j)+\pi/2}{\kappa_j},
\ea
\eeq
where
\beq\label{def_phi_j}
\ba{l}
\cos\phi_j= \sqrt{1+\frac{\eta}{|a|^2}}\,\sin \left(\frac{\kappa_j}{2}\right), 
\ea
\eeq
and 
\beq\label{def_alpha_beta}
\alpha_j:=\overline{c_j}e^{-i\eta\phi_j}-c_{-j}e^{i\eta\phi_j}, \ \ \beta_j=\overline{c_{-j}}e^{i\eta\phi_j}-c_je^{-i\eta\phi_j}.
\eeq

To construct the solution, to leading order and in terms of elementary functions, we use the matched asymptotic expansion technique introduced in \cite{GS3} for the focusing NLS model.

\subsection{The $AL_+$ case}

We first concentrate on the case $\eta=1$, giving rise to a recurrence of regular periodic AWs, in the case of one and two unstable modes.  
The instability condition $|\kappa |<\kappa_a$ implies that the first $N\le p$ modes $\pm\kappa_j,~1\le j \le N$ are unstable, where
\beq
N:=\left \lfloor{\frac{M\kappa_a}{2\pi}}\right \rfloor ,
\eeq 
and $\left \lfloor{x}\right \rfloor$ is the largest integer less or equal to $x$.

From equations (\ref{monochr_pert}), (\ref{sol_lin}), and (\ref{def_X+_X-}) it follows that the solution of the Cauchy problem (\ref{Cauchy}), for $|t|\le O(1)$, reads as follows:
\begin{equation}\label{sol_lin2}
\ba{l}
u_n(t)=a \, e^{2i|a|^2t}\bigg\{1+\eps \sum\limits_{j=1}^{N}\bigg[\frac{|\alpha_j|}{\sin(2\phi_j)}e^{\sigma_j\,t+i\phi_j}\cos\left(\kappa_j(n-X_j^+)\right) \\
+\frac{|\beta_j|}{\sin(2\phi_j)}e^{-\sigma_j\,t-i\phi_j}\cos\left(\kappa_j(n-X_j^-)\right)\bigg]+O(\eps) \mbox{ oscillations}\bigg\}+O(\epsilon^2),
\ea
\end{equation}
where $\sigma_j, \ \phi_j, \ X^{\pm}_j, \ \alpha_j, \ \beta_j, \ 1\le j\le N$ are defined respectively in \eqref{def_sigmaj_X+_X-}, \eqref{def_phi_j}, and \eqref{def_alpha_beta}. This solution grows exponentially and, when $t=O(\log(1/\eps))$, one enters the first nonlinear stage of MI.

\subsubsection{One unstable mode} In the simplest case of one unstable mode ($N=1$) only, corresponding to the case in which the period $M$ satisfies the inequalities
\beq
N=1 \ \ \ \ \Leftrightarrow \ \ \ \ \frac{2\pi}{\kappa_a} <M<\frac{4\pi}{\kappa_a},
\eeq
only the mode $\kappa_1$ is unstable, and the corresponding nonlinear stage of MI is described by the solution (\ref{Narita}) for a suitable choice of its arbitrary parameters, obtained using matched asymptotic expansions \cite{GS3}.

Matching the linearized solution (\ref{sol_lin2}) for $N=1$ and the exact solution (\ref{Narita1}) in the intermediate time interval $1\ll t \ll T_1={\sigma_1}^{-1}\log(\gamma^+/\eps)),~\gamma^+=O(1)>0$, we have
\beq
\ba{l}
u\sim ae^{2i|a|^2t}\Big[1+\eps\frac{|\alpha|}{\sin(2\phi_1)}e^{\sigma_1 t+i\phi_1}\cos\left[\kappa_1(x-X^+_1)\right]\Big],\\
{\cal N}_1\sim ae^{2i |a|^2t}e^{i(\rho-2\theta_1)}\Big[1+\eps\frac{4 G_1\cos\theta_1}{\gamma_+}e^{\sigma(K_1) t+i\theta_1}\cos\left[K_1(x-X_1)\right] \Big]
\ea
\eeq
inferring that $\rho=2\phi_1$, $K_1=\kappa_1$ (consequently $\sigma(K_1)=\sigma_1$, $\theta_1=\phi_1$), $X_1=X^+_1$, and 
\beq
\gamma^+=\frac{2\sin^2(2\phi_1)}{|\alpha_1|\cos(\kappa_1/2)} \ \ \Rightarrow \ \ T_1=t^{(1)}:=\frac{1}{\sigma_1}\log\left(\frac{\sigma^2_1}{2\eps |a|^4  |\alpha_1|\cos(\kappa_1/2)}\right).
\eeq

\begin{figure}[H]
\centering
\includegraphics[trim=0cm 0cm 0cm 0 ,width=15cm,height=10cm]{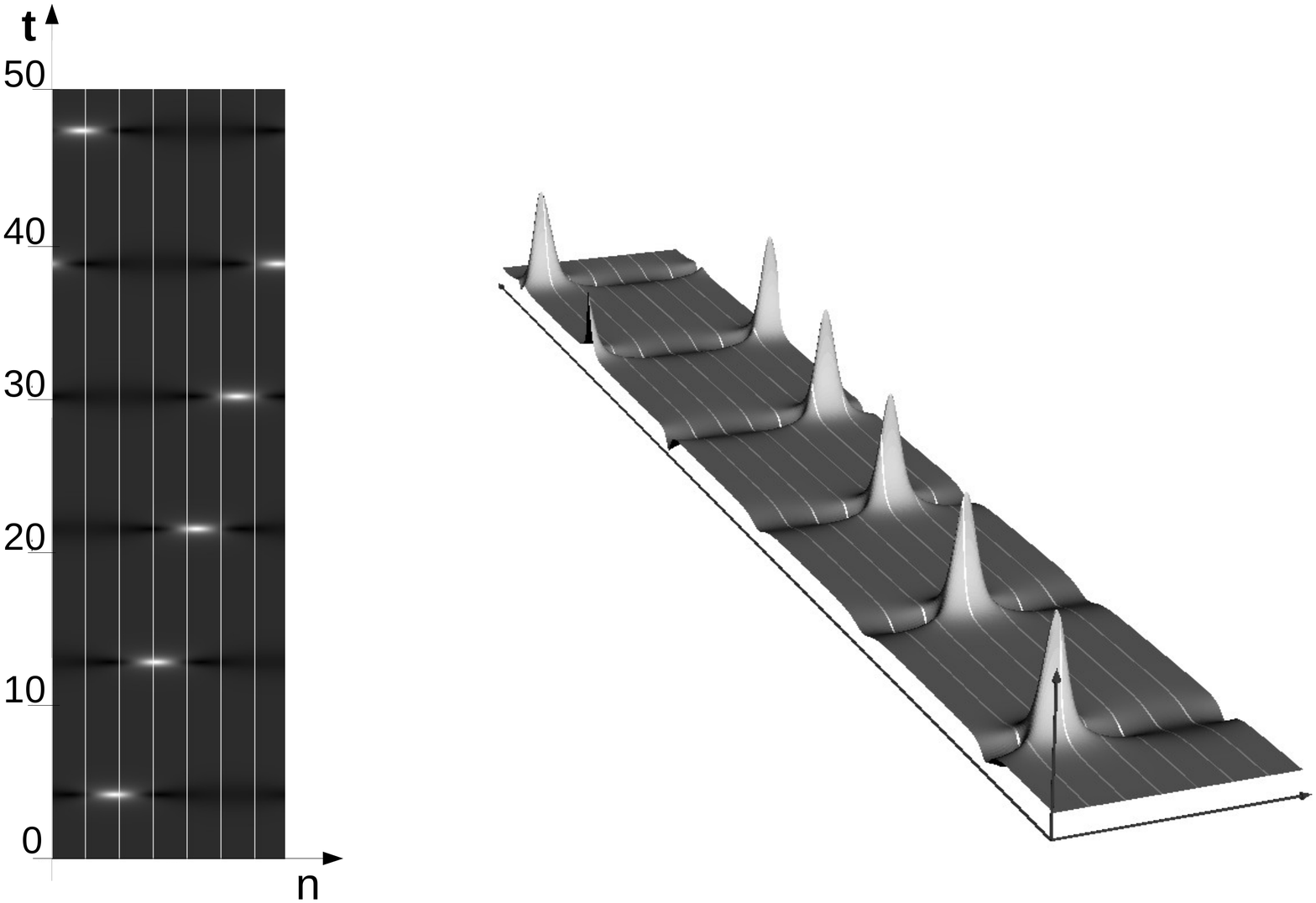}
\caption{Density and 3D plots of $|u_n(t)|$ coming from the numerical integration of the Cauchy problem of the AWs for the $AL_+$ equation ($\eta=1$), in the case of a single unstable mode. We used the 6th order Runge-Kutta method \cite{RungeKutta} with the initial condition $u_n(0)=a(1+\epsilon(c_+e^{ikn}+c_-e^{-ikn}))$, where $M=7$, $a=1.1$, $\epsilon=10^{-4}$, $c_+=0.53-i~0.86$ and $c_-=-0.26+i~0.22$. The numerical output is in perfect quantitative agreement with the theory described by \eqref{def_Delta x_Delta t},\eqref{def_recurrence}}\label{fig:recurrence1} 
\end{figure}

It follows that the Narita solution ${\cal N}_1(x,t,\kappa_1,X^+_1,t^{(1)},2\phi_1,1)$ describes the first appearance of the AW. To describe the recurrence of AWs, it is convenient to obtain the first appearance for negative times \cite{GS3}, matching the linearized solution (\ref{sol_lin2}) for $N=1$ and the Narita solution (\ref{Narita1}) in the time interval $1\ll |t| \ll T_1={\sigma_1}^{-1}\log(\gamma^-/\eps)),~\gamma^-=O(1)>0,~t<0$:
\beq
\ba{l}
u\sim ae^{2i\eta |a|^2t}\Big[1+\eps\frac{|\beta_1|}{\sin(2\phi_1)}e^{-\sigma_1 t-i\phi_1}\cos\left[\kappa_1(x-X^-_1)\right]\Big],\\
{\cal N}_1\sim ae^{2i |a|^2t}e^{i(\rho+2\theta_1)}\Big[1+\eps\frac{4 G_1\cos\theta_1}{\gamma_-}e^{-\sigma(K_1) t-i\theta_1}\cos\left[K_1(x-X_1)\right] \Big],
\ea
\eeq
Comparison gives $\rho=-2\phi_1$, $K_1=\kappa_1$ (consequently $\sigma(K_1)=\sigma_1$, $\theta_1=\phi_1$), $X=X^-_1$,  and 
\beq
\gamma^-=\frac{2\sin^2(2\phi_1)}{|\beta_1|\cos(\kappa_1 /2)} \ \ \Rightarrow \ \ T_1=t^{(0)}:=-\frac{1}{\sigma_1}\log\left(\frac{\sigma^2_1}{2 |a|^4 \eps |\beta_1|\cos(\kappa_1/2)}\right).
\eeq
It follows that the solution ${\cal N}_1(x,t,\kappa_1,X^-_1,t^{(0)},-2\phi_1,1)$ describes the first appearance of the AW also at negative times. Comparing the two consecutive appearances at times $t^{(0)}$ and $t^{(1)}$,  and using the time translation symmetry of $AL_{+}$, we infer the following periodicity law for the general recurrence of the $AL_+$ AWs (see \cite{GS3} for more details)
\beq
u(x+\Delta x,t+\Delta t)=u(x,t)+O(\eps),
\eeq
where
\beq\label{def_Delta x_Delta t}
\ba{l}
\Delta x=X^+_1-X^-_1=\frac{\arg(\alpha_1\beta_1)}{\kappa_1}, \\
\Delta t=t^{(1)}-t^{(0)}=\frac{1}{\sigma_1}\log\left(\frac{\sigma^4_1}{4 \eps^2 |a|^8 |\alpha_1\beta_1|\cos^2(\kappa_1/2)}\right).
\ea
\eeq
Summarizing, the $n^{th}$ AW appearance in the FPUT recurrence generated by the Cauchy problem (\ref{Cauchy}), in the case of the single unstable mode $\kappa_1$, is described, in the time interval $|t-t^{(n)}|=O(1)$, by the Narita solution ${\cal N}_1(x,t;\kappa_1,x^{(n)},t^{(n)},1,\rho_n)$ up to $O(\eps)$ errors, where
\beq\label{def_recurrence}
\ba{l}
x^{(n)}=x^{(1)}+(n-1)\Delta x, \ \ x^{(1)}=\frac{\arg(\alpha_1)+\pi/2}{\kappa_1}, \ \ \mod M, \\
t^{(n)}=t^{(1)}+(n-1)\Delta t, \ \ t^{(1)}=\frac{1}{\sigma_1}\log\left(\frac{\sigma^2_1}{2 \eps |a|^4 |\alpha_1|\cos(\kappa_1/2)}\right),\\
\rho_n=2\phi_1+(n-1)4\phi_1,
\ea
\eeq
and $\Delta x$ and $\Delta t$ are defined in (\ref{def_Delta x_Delta t}). This is the analytic and quantitative description of the FPUT recurrence of AWs of the $AL_+$ equation in term of the initial data through elementary functions. $x^{(1)}$ and $t^{(1)}$ are respectively the first appearance time and the position of the maximum of the absolute value of the AW; $\Delta x$ is the $x$-shift of the position of the maximum between two consecutive appearances, and  $\Delta t$ is the time interval between two consecutive appearances (see Figure~\ref{fig:recurrence1}).
The quantitative agreement between the above theory and numerical experiments is perfect, as one can see from the following table, in which one compares the  values $(x^{(j)},t^{(j)})$ of the position and time of the $j^{th}$ appearance of the AW, for $j=1,\dots,6$, as predicted by \eqref{def_Delta x_Delta t}-\eqref{def_recurrence} with the values coming from the numerical experiment of Figure~\ref{fig:recurrence1}. E.g., for $\eps=10^{-4}$, the first disagreement in the $6^{th}$ appearance is in the $7^{th}$ decimal digit, corresponding to the $9^{th}$ significant digit!  
\begin{table}[h!]
\begin{center}
\begin{tabular}{l l r}
\hline
\hline
\hspace{5mm} &    Numeric & Theor  \\
\hline

($x^{(1)}$, $t^{(1)}$) & (1.87977083,$\;$   4.17892477) &    (1.87977074, $\;$  4.17892429)\\
($x^{(2)}$, $t^{(2)}$) & (3.08443357,$\;$   12.8077003) &    (3.08443336, $\;$  12.8076998)\\
($x^{(3)}$, $t^{(3)}$) & (4.28909633,$\;$   21.43647586) &   (4.28909597, $\;$  21.43647549 )\\
($x^{(4)}$, $t^{(4)}$) & (5.49375910,$\;$   30.06525140) &   (5.49375859, $\;$  30.06525108)\\
($x^{(5)}$, $t^{(5)}$) & (6.69842186 ,$\;$  38.69402694) &   (6.69842121, $\;$  38.69402668)\\
($x^{(6)}$, $t^{(6)}$) & (0.90308462,$\;$   47.32280249) &   (0.90308383, $\;$  47.32280228 )\\
\hline
\end{tabular}
\end{center}
\end{table}
\newpage

In addition, using the fact that, at each appearance, the AW is exponentially localized in time, and that, from (\ref{phase_shift}), after each appearance, the background exhibits a $4\phi_1$ phase shift, the FPUT recurrence can be described by the following expression, uniform in space-time, with $t\le t^{(n)}+O(1)$:
\beq\label{FPUT_AL}
u(x,t)=\sum\limits_{j=0}^n{\cal N}_1(x,t;\kappa_1,x^{(j)},t^{(j)},\rho_j,1)-ae^{2i|a|^2t}\frac{1-e^{4i\phi_1 n}}{1-e^{4i\phi_1}}+O(\eps).
\eeq

We remark, from (\ref{def_recurrence}), that the maximum of the AW at the first appearance is located on a lattice point $x^{(1)}\in\ZZ$, if the initial data are such that
\beq
\frac{\arg\alpha_1}{\kappa_1}+\frac{N}{4}\in\ZZ, 
\eeq
If, in addition, $\Delta x\in\ZZ$, i.e., from (\ref{def_recurrence}), (\ref{def_Delta x_Delta t})
\beq
\frac{\arg\beta_1}{\kappa_1}-\frac{N}{4}\in\ZZ,
\eeq
then the maxima of the FPUT recurrence are all located on the lattice points.\\
\ \\
{\bf The distinguished case $\alpha_1\beta_1\in\RR$}. As in the NLS case, a very distinguished situation occurs when the initial data are such that
\beq\label{alpha_beta_real}
\alpha_1\beta_1\in\RR .
\eeq
Indeed, from \eqref{def_alpha_beta},\eqref{def_Delta x_Delta t}:
\begin{itemize}
\item If $\alpha_1\beta_1>0$, then $\Delta x=0$ and the FPUT recurrence is periodic with period $\Delta t$.
\item If $\alpha_1\beta_1<0$, then $\Delta x=M/2$ and the FPUT recurrence is periodic with period $2 \Delta t$ .
\end{itemize}

It is easy to verify that
\beq
\alpha_1\beta_1\in\RR \ \ \ \Leftrightarrow \ \ \ |c_1|=|c_{-1}|=:|c|,
\eeq
with
\beq
\alpha_1\beta_1 =2 |c|^2\left[\cos\gamma -\cos(2\phi_1) \right], \ \ \gamma:=\arg(c_1)+\arg(c_{-1}),
\eeq
Therefore, in terms of the initial data :
\begin{itemize}
\item   $|c_1|=|c_{-1}|, \ |\gamma|<2\phi_1 \ \ \ \Leftrightarrow \ \ \ \alpha_1\beta_1>0$,
\item $|c_1|=|c_{-1}|, \ |\gamma|>2\phi_1 \ \ \ \Leftrightarrow \ \ \ \alpha_1\beta_1<0$,
\end{itemize}
Particularly interesting subcases are $u_n(0)\in\RR$ and $u_n(0)\in i\RR$. \\
a) If $u_n(0)\in\RR$, then $|c_1|=|c_{-1}|, \ \gamma=0$, implying $\alpha_1\beta_1>0$, $\Delta x=0$, and a periodic FPUT recurrence with period $\Delta t$.\\
b) If $u_n(0)\in i\RR$, then $|c_1|=|c_{-1}|, \ |\gamma|=\pi$, implying $\alpha_1\beta_1<0$, $\Delta x=M/2$, and a periodic FPUT recurrence with period $2 \Delta t$.

We remark that the condition (\ref{alpha_beta_real}) is not generic with respect to the AL dynamics, since it arises imposing the real constraint $|c_1|=|c_{-1}|$ on the initial data. But, as we shall see in the forthcoming paper \cite{CS3}, it becomes the generic asymptotic state when the $AL_+$ dynamics is perturbed by a small loss or gain.

\subsubsection{Two unstable modes} 

In the  case of two unstable modes ($N=2$), corresponding to the case in which the period $M$ satisfies the inequalities
\beq
N=2 \ \ \ \ \Leftrightarrow \ \ \ \ \frac{4\pi}{\kappa_a} <M<\frac{6\pi}{\kappa_a},
\eeq
only the modes $\kappa_1$  and $\kappa_2$ are unstable, and the corresponding nonlinear stage of MI is described by the solution ${\cal N}_2$ for a suitable choice of its arbitrary parameters.

Matching the linearized solution (\ref{sol_lin2}) for $N=2$ and the $AL_+$ solution \eqref{Narita2} in the intermediate time interval $1\ll t \ll 2\left({\sigma_1}^{-1}+{\sigma_2}^{-1}\right)|\log(\eps))|$, we have
\begin{equation}
\ba{l}
u\sim a\,e^{2i|a|^2t}\left[1+\eps\sum\limits_{j=1}^2\left(\frac{|\alpha_j|}{\sin(2\phi_j)}e^{\sigma_j t+i\phi_j}\cos\left[\kappa_j(n-X_j^+)\right]\right)\right],\\
{\cal N}_2\sim a\,e^{2i |a|^2t}e^{i(\rho-2 (\theta_1+\theta_2))}\left[1+ \sum\limits_{j=1}^2\left(\frac{2 a_{12}\sin(2\theta_j)}{\cos\left(\frac{K_j}{2}\right)}e^{\Sigma_j (t-T_j)+i \theta_j}\cos\left[K_j(n-X_j)\right]\right)\right],
\ea
\end{equation}
inferring that $\rho=2(\theta_1+\theta_2)$, $K_j=\kappa_j$ (consequently $\theta_j=\phi_j$ and $\Sigma_j=\sigma_j$), $X_i=X_i^+$, for $i=1,2$, and  
\begin{equation}\label{def:t1+_t2+}
\begin{split}
&T_j=t_j^+:=\frac{1}{\sigma_j}\log\left(\frac{a_{12}\;\sigma^2_j}{2 \eps |a|^4 |\alpha_j|\cos(\kappa_j/2)}\right), \ \ j=1,2.
\end{split}
\end{equation}
It follows that the solution $ {\cal N}_2(n,t;\kappa_1,\kappa_2,X_1^+,X^+_2,t_1^+,t_2^+,2(\theta_1+\theta_2),1)$ describes the first appearance of the AW.

\begin{figure*}[h!!]
  \begin{center}
    \includegraphics[height=9 cm,width= 13 cm]{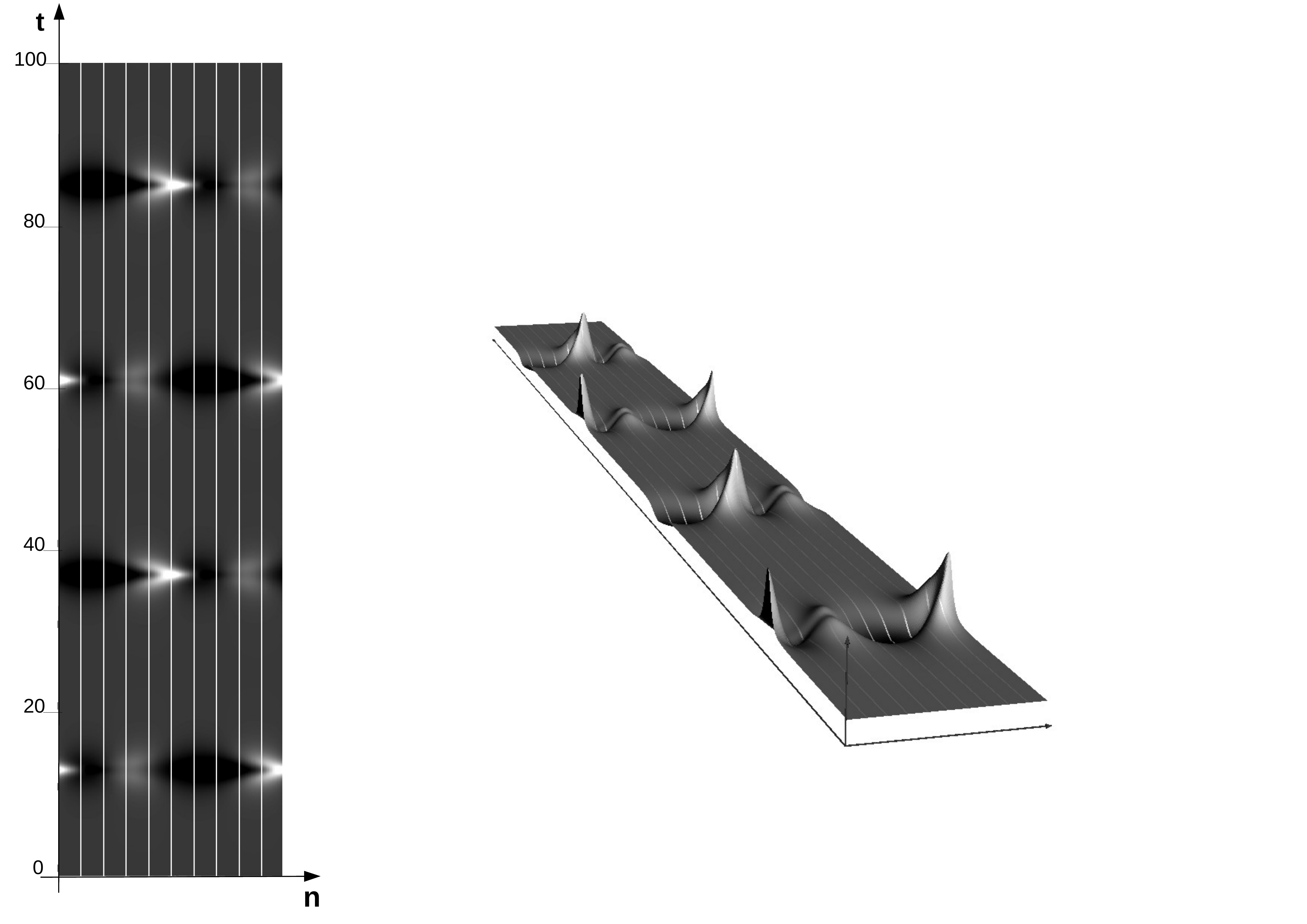} \caption{Density and 3D plots of $|u(n,t)|$, with two unstable modes, with $M=10$, $a=0.8$ and$ \eps=10^{-5}$. The numerical integration is performed using the 6th order Runge-Kutta. The initial condition is $u_n(0)=a(1+\eps(c_{1}e^{i\kappa_1 n}+c_{-1}e^{-i\kappa_1 n}+c_{2}e^{i\kappa_2 n}+c_{-2}e^{-i\kappa_2 n}))$, where $\kappa_1=\frac{2\pi}{M}$, $\kappa_2=2\kappa_1$, $c_{1}=-0.2541+i~0.6967$, $c_{-1}=-0.3492+i~0.6642$, $c_{2}=1.016+i~16.87$, $c_{-2}=-6.402-i~15.19$, and $\eps=10^{-5}$. The coefficients $c_{\pm j}$ are chosen to have $|t_1^{(j)}-t_2^{(j)}|\ll 1, \ j=0,1$; these two conditions imply that the two unstable modes appear approximately at the same time for many recurrences. \label{recurrence2}}
    \end{center}
\end{figure*}

Proceeding as in the case of a single unstable mode, we describe the first appearance for negative times, matching the linearized solution (\ref{sol_lin}) for $N=2$ and the solution (\ref{Narita2}) in the time interval $1\ll |t| \ll \left({\sigma_1}^{-1}+{\sigma_2}^{-1}\right)|\log( \eps))|, \ t<0$:
\begin{equation}
\ba{l}
u\sim ae^{2i|a|^2t}\left[1+\eps\sum\limits_{j=1}^2\left(\frac{|\beta_j|}{\sin(2\phi_j)}e^{-\sigma_j t-i\phi_j}\cos\left[\kappa_j(x-X_j^-)\right]\right)\right], \\
{\cal N}_2\sim a\,e^{2i |a|^2t}e^{i(\rho+2 (\theta_1+\theta_2))}\left[1+ \sum\limits_{j=1}^2\left(\frac{2 a_{12}\sin(2\theta_j)}{\cos\left(\frac{K_j}{2}\right)}e^{-\Sigma_j (t-T_j)-i \theta_j}\cos\left[K_j(n-X_j)\right]\right)\right], 
\ea
\end{equation}
inferring that $\rho=-2(\theta_1+\theta_2)$, $K_j=\kappa_j$ (consequently $\theta_j=\phi_j$ and $\Sigma_j=\sigma_j$), $X_j=X^-_j$,  and 
\begin{equation}
\begin{split}
&T_j=t_j^-:=-\frac{1}{\sigma_j}\log\left(\frac{a_{12}\;\sigma^2_j}{2 \eps |a|^4 |\beta_j|\cos(\kappa_j/2)}\right), \ \ j=1,2.
\end{split}
\end{equation}
It follows that the solution $ {\cal N}_2(n,t;\kappa_1,\kappa_2,X_1^-,X^-_2,t_1^-,t_2^-,1,-2(\theta_1+\theta_2),1)$ describes the first appearance of the AW at negative times. As in the one mode case, comparing the two consecutive appearances  we construct the solution of the Cauchy problem to leading order. Introduce:
\beq\label{def_Delta x_Delta t_2}
\ba{l}
\Delta x_j=X^+_j-X^-_j=\frac{\arg(\alpha_j\beta_j)}{\kappa_j}, \ \ \ \ j=1,2, \\[2mm]
\Delta t_j=t^+_j-t^-_j=\frac{2}{\sigma_j}\log\left(\frac{\sigma^2_j}{2 |a|^4 \eps \sqrt{|\alpha_j\beta_j|}\cos(\kappa_j/2)}\right).
\ea
\eeq
Then the $n^{th}$ AW appearance in the FPUT recurrence generated by the Cauchy problem, in the case of two unstable modes $\kappa_1$ and $\kappa_2$, is described, in the time interval $|t-\left(t^{(n)}_1+t^{(n)}_1\right)/2|=O(1)$, by the solution\\ $ {\cal N}_2\big(n,t;\kappa_1,\kappa_2,x^{(n)}_1,x^{(n)}_2,t^{(n)}_1,t^{(n)}_2,-2(\theta_1+\theta_2),1\big)$ up to $O(\eps)$ errors, where
\beq\label{def_recurrence_2}
\ba{l}
x^{(n)}_j=x^{(1)}_j+(n-1)\Delta x_j, \ \ \ x^{(1)}_j=X_j^+=\frac{\arg(\alpha_j)+\pi/2}{\kappa_j}, \ \ \mod M, \\[2mm]
t^{(n)}_j=t^{(1)}_j+(n-1)\Delta t_j, \ \ \ t^{(1)}_j=t_j^+=\frac{1}{\sigma_j}\log\left(\frac{\sigma^2_j}{2 |a|^2 \eps |\alpha_j|\cos(\kappa_j/2)}\right),\\[2mm]
\rho^{(n)}=2(\phi_1+\phi_2)+4(n-1)(\phi_1+\phi_2),
\ea
\eeq
and $\Delta x_j$ and $\Delta t_j$ are defined in (\ref{def_Delta x_Delta t}). This is the analytic and quantitative description of the FPUT recurrence of AWs of the AL equation in term of the initial data through elementary functions (see Figures \ref{recurrence2}).

Equivalently, using the fact that, at each appearance, the AW is exponentially localized in time and the background exhibits a $4(\phi_1+\phi_2)$ phase shift, the FPUT recurrence can be described by the following expression, uniform in space-time, with $t\le \left(t^{(n)}_1+t^{(n)}_1\right)/2+O(1)$:
\beq\label{FPUT_AL}
\ba{l}
u(x,t)=\sum\limits_{j=0}^n {\cal N}_2(n,t;\kappa_1,\kappa_2,n_1^{(j)},n_2^{(j)},t_1^{(j)},t_2^{(j)},\rho^{(j)}) \\
-\,a\,e^{2i|a|^2t}\frac{1-e^{4i(\phi_1+\phi_2) n}}{1-e^{4i(\phi_1+\phi_2)}}+O(\eps).
\ea
\eeq
It is important to remark that the recurrence results of this subsection for the case of two unstable modes are valid if the
initial data are such that the first appearance times of the two unstable modes, for positive and negative $t$, are approximately the same: $|t_1^{(j)}-t_2^{(j)}|\ll 1, \ j=0,1$, implying that the unstable modes appear approximately at the same time for many recurrences (see Figure \ref{recurrence2}). If the appearance times of the two modes are sensibly different, the picture is more complicated and the finite-gap approach is the proper tool to analyze it, as it was done in \cite{GS2} for the NLS model.

\subsection{The $AL_-$  case}

If $\eta=-1$ and $|a|>1$, all modes $\pm\kappa_j,~1\le j \le p$ are unstable, and the solution of the Cauchy problem (\ref{Cauchy}), for $|t|\le O(1)$, reads as follows:
\begin{equation}\label{sol_lin_def}
\ba{l}
u_n(t)=a \, e^{-2i|a|^2t}\bigg\{1-\eps \sum\limits_{j=1}^{p}\bigg[\frac{|\alpha_j|}{\sin(2\phi_j)}e^{\sigma_j\,t-i\phi_j}\cos\left(\kappa_j(n-X_j^+)\right) \\
+\frac{|\beta_j|}{\sin(2\phi_j)}e^{-\sigma_j\,t+i\phi_j}\cos\left(\kappa_j(n-X_j^-)\right)\bigg]+O(\epsilon^2),
\ea
\end{equation}
where $\sigma_j, \ \phi_j, \ X^{\pm}_j, \ \alpha_j, \ \beta_j, \ 1\le j\le p$ are defined respectively in \eqref{def_sigmaj_X+_X-}, \eqref{def_phi_j}, and \eqref{def_alpha_beta}.

This solution grows exponentially and generically develops singularities of the type discussed in \S 3 in the first nonlinear stage, when $t=O(\log(1/\eps))$. Therefore it does no make sense to study its recurrence properties, but it does make sense to see how a smooth initial condition  \eqref{Cauchy}     evolves into a singularity in finite time.

If $p=1$, then the period is $M=3$, and $\kappa_1$ is the only unstable mode. The matching procedure of the previous section leads to the comparison between the linearized solution \eqref{sol_lin_def} and the one mode solution \eqref{Narita} in the intermediate region $O(1)\ll t \ll \sigma_1^{-1}\log(|\eps|)$:
\beq
\ba{l}
u\sim a e^{-2i|a|^2t}\Big[1-\eps\frac{|\alpha_1|}{\sin(2\phi_1)}e^{\sigma_1 t-i\phi_1}\cos\left[\kappa_1(x-X^+_1)\right]\Big],\\
{\cal N}_1\sim a e^{-2i |a|^2t}e^{i(\rho+2\phi_1)}\Big[1-\eps\frac{4 G_1\cos\theta_1}{\gamma^+}e^{\Sigma_1 (t-T_1)-i\theta_1}\cos\left[K_1(x-X_1)\right] \Big], \\
T_1=\frac{1}{\sigma_1}\log\left(\frac{\gamma^+}{\eps} \right),
\ea
\eeq
inferring that $\rho=-2\phi_1$,  $\kappa=\kappa_1$ (consequently $\theta_1=\phi_1$ and $\Sigma_1=\sigma_1$), $X_1=X^+_1$, and 
\beq
\gamma^+=\frac{2\sin^2(2\phi_1)}{|\alpha_1|\cos(\kappa_1/2)} \ \ \Rightarrow \ \ T_1=t^{(1)}:=\frac{1}{\sigma_1}\log\left(\frac{\sigma^2_1}{2 \eps  |a|^4|\alpha_1|\cos(\kappa_1/2)}\right).
\eeq
Therefore the smooth initial condition $u_n(0)=a\left[1+\eps\left(c_1 e^{i\kappa_1 n}+c_{-1} e^{-i\kappa_1 n} \right) \right]$ evolves into the generically singular solution
\beq
u_n(t)\sim {\cal N}_1\left(n,t;\kappa_1, X^+_1,t^{(1)},-2\phi_1,-1 \right), \ \ |t-t^{(1)}|=O(1),
\eeq
whose singularity properties have been studied in \S 3. Analogously, in the case $p=2$, corresponding to the two unstable modes $\kappa_j, \ j=1,2$ and to the period $M=5$, the smooth initial condition
\beq
u_n(0)=a\left[1+\eps\sum\limits_{j=1}^2\left(c_j e^{i\kappa_j n}+c_{-j} e^{-i\kappa_j n} \right) \right]
\eeq
evolves into the generically singular two-breather solution 
\beq
u_n(t)\sim {\cal N}_2\left(n,t;\kappa_1,\kappa_2,X^+_1,X^+_2,t^{(1)}_1,t^{(1)}_2,-2(\phi_1+\phi_2),-1 \right)
\eeq
(see Figure \ref{singg}).

\section{Appendix. Darboux transformations and periodic AW solutions}

We look for a gauge transformation matrix $D_n(t,\lambda)$ (the so-called Darboux matrix) that preserves the structure and the symmetries of the AL Lax pair. More precisely, let $(u_n^{[0]},{\vec\psi}_n^{[0]})$ and $(u_n,{\vec\psi}_n)$ be solutions of the Lax pair \eqref{AL_Lax_1}. Then we look for the transformation
\beq\label{psi_gauge}
{\vec\psi}_n=D_n(t,\lambda){\vec\psi}_n^{[0]},
\eeq
implying the following two equations for the Darboux matrix
\beq\label{L_T_gauge}
D_{n+1}L_n^{[0]}=L_n D_{n}, \ \ \ \ {D_n}_t=A_n D_n -D_n A_n^{[0]},
\eeq
where $(L_n^{[0]},A_n^{[0]})$ are the matrices $(L_n,A_n)$ in \eqref{AL_Lax_1} in which $u_n$ is replaced by $u_n^{[0]}$, together with the symmetry 
\begin{equation}\label{darb_sym}
D_n(\lambda)=P_\eta\, \overline{D_n\left(\frac{1}{\overline{\lambda}}\right)}P_\eta^\dagger
\end{equation}\\
coming from \eqref{symmetries},\eqref{def_P}.

Following \cite{Geng}, we look for the Darboux matrix in the form:\\[2mm]
\begin{equation}\label{matrix_form}
D_n^{(N)}(\lambda)=\begin{pmatrix}
\lambda^N+\sum\limits_{l=1}^{N}a_n^{(N-2\,l)}\lambda^{N-2\,l}\;\;\;\;& \sum\limits_{l=1}^{N}b_n^{(N-2\,l+1)}\lambda^{N-2\,l+1}\\[6mm]
\sum\limits_{l=1}^{N}c_n^{(N-2\,l+1)}\lambda^{-N+2\,l-1} &\lambda^{-N}+\sum\limits_{l=1}^{N}d_n^{(N-2\,l)}\lambda^{-N+2\,l}
\end{pmatrix},
\end{equation}
for $N\in\NN^+$, corresponding to the Darboux transformation
\beq
{\vec\psi}_n^{[N]}=D_n(t,\lambda){\vec\psi}_n^{[0]}; 
\eeq
then the  symmetry (\ref{darb_sym}) implies the following relations between the matrix elements of $D_n^{(N)}$:
\beq\label{def_c_d}
d_n^{(l)}=\overline{a_n^{(l)}}, \ \ \ c_n^{(l)}=-\eta\overline{b_n^{(l)}}.
\eeq

It is well-known that the Darboux matrix describes a one-parameter (the complex parameter $\lambda$) family of transformations becoming singular at one or more values of $\lambda$. If $\lambda_i$ is a singular point such that $\det D_n(\lambda_i,t)=0$, the symmetries of the Lax pair imply that also $-\lambda_i$ and $\pm 1/\overline{\lambda}_i$ are singular points. At the singular points ($\pm\lambda_i$, $\pm 1/\overline{\lambda}_i$) the matrix $D_n(\lambda,t)$ has range 1 and, if $\underline{\mathbb{\xi}}_n(t,\lambda),\underline{\mathbb{\chi}}_n(t,\lambda)$ are the columns of the fundamental matrix solution $\Psi_n^{[0]}(t,\lambda)$ of the Lax pair \eqref{AL_Lax_1} for $u_n=u_n^{[0]}$: $\Psi_n^{[0]}(t,\lambda) =\left(\underline{\mathbb{\xi}}_n(t,\lambda),\underline{\mathbb{\chi}}_n(t,\lambda)\right)$, their images must be proportional in the points  ($\pm\lambda_i$, $\pm 1/\overline{\lambda_i}$) where the matrix is singular: \\
\beq\label{degeneracy2}
D_n(\lambda_i)\cdot\left(\underline{\xi}_n(\lambda_i)-\gamma_i\underline{\chi}_n(\lambda_i)\right)=\underline{0},
\eeq
where $\gamma_i$ is the proportionality factor.

Equation (\ref{degeneracy2}) implies, for each $\lambda_i$, the system:
\begin{equation}\label{ab_sys}
\begin{split}
&\left(\lambda_i^N+\sum\limits_{l=1}^{N}a_n^{(N-2\,l)}\lambda^{N-2\,l}_i\right)+ \left(\sum\limits_{l=1}^{N}b_n^{(N-2\,l+1)}\lambda_i^{N-2\,l+1}\right)r_i=0, \\[2mm]
&\left(\frac{1}{\overline{\lambda}_i^N}+\sum\limits_{l=1}^{N}\frac{a_n^{(N-2\,l)}}{\overline{\lambda}^{N-2\,l}_i}\right)-\eta \left(\sum\limits_{l=1}^{N}\frac{b_n^{(N-2\,l+1)}}{\overline{\lambda}_i^{N-2\,l+1}}\right)\frac{1}{\overline{r_i}}=0, \\
\end{split} 
\end{equation}\\[2mm]
where
\begin{equation}\label{r_i}
r_i=\frac{\left({\xi}_n(\lambda_i)\right)_2-\gamma_i\;\left({\chi}_n(\lambda_i)\right)_2}{\left({\xi}_n(\lambda_i)\right)_1-\gamma_i\;\left({\chi}_n(\lambda_i)\right)_1}.
\end{equation}
Note that a change of the basis of the eigenvectors $\left(\underline{\mathbb{\xi}}_n(t,\lambda),\underline{\mathbb{\chi}}_n(t,\lambda)\right)$ is equivalent to a rescaling of the constant $\gamma_i$. If the number of singular points $\lambda_i$ is equal to the order $N$ of the Darboux transformation, the relations \eqref{ab_sys} for $i=1,\dots,N$ define a determined system of $2N$ equations for the $2N$ unknowns $a_n^{(l)}, \ b_n^{(l)}$, $l=1,\dots, N$, that can be uniquely solved, leading to the wanted Darboux matrix.

At last, from the the first of equations (\ref{L_T_gauge}), the dressed solution $u^{[N]}_n(t)$ can be calculated from the solution $u^{[0]}_n(t)$ in the following way:\\ 
\begin{equation}\label{dress}
u^{[N]}_n(t)=u^{[0]}_n(t) a_{n+1}^{(-N)}(t)+b_{n+1}^{(-N+1)}(t).
\end{equation}

Now we specialize the previous formulas for the two simplest cases $N=1,2$.
\subsubsection*{N=1}
If $N=1$, the linear system \eqref{ab_sys} of two equations yields the solution\\
\beq\label{def_a1_b1}
\ba{l}
 a_n^{(-1)}(t)=\frac{\Delta_a^{(1)}}{\Delta^{(1)}}=-\frac{\eta\lambda_1+\dfrac{|r_1|^2}{\overline{\lambda_1}}}{|r_1|^2\overline{\lambda_1}+\frac{\eta}{\lambda_1}}, \\[2mm]
 b_n^{(0)}(t)=\frac{\Delta_b^{(1)}}{\Delta^{(1)}}=-\frac{\overline{r_1}}{|r_1|^2\overline{\lambda_1}+\frac{\eta}{\lambda_1}}\left(|\lambda_1|^2-\frac{1}{|\lambda_1|^2}\right),
 \ea
\eeq
 where 
 \begin{equation*}
\Delta_a^{(1)}=\det\begin{pmatrix}{-}\lambda_1&&r_1\\[2mm]{-}\dfrac{1}{\overline{\lambda_1}}&&-\dfrac{\eta}{\overline{r_1}}\end{pmatrix},\hspace{2mm}\Delta_b^{(1)}=\det\begin{pmatrix}\frac{1}{\lambda_1}&&{-}\lambda_1\\[2mm]\overline{\lambda_1}&&{-}\dfrac{1}{\overline{\lambda_1}}\end{pmatrix},\hspace{2mm}\Delta^{(1)}=\det\begin{pmatrix}\frac{1}{\lambda_1}&&r_1\\[2mm]\overline{\lambda_1}&&-\dfrac{\eta}{\overline{r_1}}\end{pmatrix}.
\end{equation*}

At last: 
 \begin{equation}\label{1sol}
u^{[1]}_n(t)=u^{[0]}_n(t) a_{n+1}^{(-1)}(t)+b_{n+1}^{(0)}(t).
\end{equation}

\subsubsection*{N=2}
If $N=2$, we can write the relevant Darboux matrix elements $a^{(-2)}_n$ and $b^{(1)}_n$ through the formulas:
\beq\label{def_a2_b2}
a_n^{(-2)}(t)=\frac{\Delta^{(2)}_a}{\Delta^{(2)}},\hspace{1cm}b_n^{(-1)}(t)=\frac{\Delta^{(2)}_b}{\Delta^{(2)}},
\eeq
where
\begin{equation*}
\Delta^{(2)}=\det\begin{pmatrix}
1&&\frac{1}{\lambda_1^2}&&r_1\lambda_1&&\frac{r_1}{\lambda_1}\\[2mm]
1&&\frac{1}{\lambda_2^2}&&r_2\lambda_2&&\frac{r_2}{{\lambda_2}}\\[2mm]
1&&\overline{\lambda_1}^2&&-\dfrac{\eta}{\overline{\lambda}_1\overline{r_1}}&&-\dfrac{\eta\overline{\lambda}_1}{\overline{r_1}}\\[2mm]
1&&\overline{\lambda_2}^2&&-\dfrac{\eta}{\overline{\lambda}_2\overline{r_2}}&&-\dfrac{\eta\overline{\lambda}_2}{\overline{r_2}}
\end{pmatrix},
\end{equation*}
and $\Delta_a^{(2)}$ and $\Delta_b^{(2)}$ are obtained substituting in $\Delta^{(2)}$ the second and the fourth columns, respectively, by the vector $(-\lambda_1^2,-\lambda_2^2,-{1}/{\overline{\lambda_1}^2},-{1}/{\overline{\lambda_2}^2})^T$. As for the previous case, specializing \eqref{dress} for $N=2$, we obtain the solution:
\begin{equation}\label{2sol}
u^{[2]}_n(t)=u^{[0]}_n(t) a_{n+1}^{(-2)}(t)+b_{n+1}^{(-1)}(t).
\end{equation}

\subsubsection*{The Darboux dressing of the background solution}

Now we specialize this construction choosing $u^{[0]}_n(t)=a \,e^{2i\eta |a|^2 t}$ (the background solution); then the matrix fundamental solution of the Lax pair  reads:
\beq\label{fund_2}
\ba{l}
\Psi^{[0]}_n(\lambda;t)=\left(\underline{\mathbb{\xi}}_n(t,\lambda),\underline{\mathbb{\chi}}_n(t,\lambda)\right)=\\
\left(1+\eta\,|a|^2\right)^{\frac{n}{2}}e^{i\left(|a|^2t+\frac{\arg a}{2}\right)\sigma_3}\begin{pmatrix}
	e^{i\frac{\eta}{2}(kn-\phi)-\frac{\sigma t}{2}}&&-e^{-i\frac{\eta}{2}(kn-\phi)+\frac{\sigma t}{2}}\\
	\frac{i}{\sqrt{\eta}}e^{i\frac{\eta}{2}(kn+\phi)-\frac{\sigma t}{2}}&&\frac{i}{\sqrt{\eta}}e^{-i\frac{\eta}{2}(kn+\phi)+\frac{\sigma t}{2}}
\end{pmatrix}e^{i\nu\,t}, \\
\ea
\eeq\\[2mm]
where
\beq\label{rel_lambda_k}
\ba{l}
\cos\left(\frac{k}{2}\right)=\frac{(\lambda+\lambda^{-1})}{2\sqrt{1+\eta|a|^2}} \ \Leftrightarrow \ \lambda=\sqrt{1+\eta |a|^2}\cos(\kappa/2)+|a|\sqrt{\eta}\sin\phi, \\
\cos\phi=\sqrt{1+\frac{\eta}{|a|^2}}\sin\left(\frac{k}{2}\right), \\
\nu=2|a|\sqrt{\eta+|a|^2}\sin\phi \cos\frac{k}{2}.
\ea
\eeq

Therefore the building blocks of the solutions  \eqref{1sol}, \eqref{2sol}, are the functions ${r_i}(n,t) $, calculated from (\ref{fund_2}) and \eqref{r_i} in the following form:
\begin{equation}\label{ridef}
{r_i}=-\sqrt{\eta}\;\dfrac{\sin\left(\dfrac{\kappa_i (n-n_i)+i\eta\sigma_i (t-t_i) +\phi_i}{2} \right)}{\cos\left(\dfrac{\kappa_i (n-n_i)+i\eta\sigma_i (t-t_i) -\phi_i}{2} \right)}\;e^{-2i\eta|a|^2 t-i { \arg a}},
\end{equation}\\[2mm]
where $n_i=\eta\frac{\arg\gamma_i}{\kappa_i}$ and  $t_i=-\frac{log(|\gamma_i|)}{\sigma_i}$, and the singularities $\lambda_i$ are expressed in terms of the modes $\kappa_i$ and the angles $\phi_i$ via \eqref{rel_lambda_k}:
\beq\label{def_lambda_i}
\lambda_i=\sqrt{1+\eta |a|^2}\cos(\kappa_i/2)+|a|\sqrt{\eta}\sin\phi_i.
\eeq
 Substituting \eqref{ridef} and \eqref{def_lambda_i} into equations \eqref{def_a1_b1} and \eqref{def_a2_b2} we construct the relevant coefficients of the Darboux matrices under construction;  then formulas \eqref{1sol} and \eqref{2sol} give the wanted solutions, equivalent to the  one and two breather solutions \eqref{Narita1} and \eqref{Narita2}, through the following relations among the parameters.\\
\ \\
For the one mode solution ${\cal N}_1(n,t;K_1,X_1,T_1,\rho,\eta)$ in \eqref{Narita1}: $K_1=\kappa_1$, $\theta_1=\phi_1$, $X_1=n_1-\frac{\eta}{2}-\frac{\pi}{2K_1}$, $T_1=t_1$, and $\rho=\pi$.\\
\ \\
For the two mode solution ${\cal N}_2(n,t;K_1,K_2,X_1,X_2,T_1,T_2,\rho,\eta)$ in \eqref{Narita2}: $K_j=\kappa_j$, $\theta_j=\phi_j$, $X_1=n_1-\frac{1}{2}-\frac{\pi}{2 K_1}$, $X_2=n_2-\frac{1}{2}+\frac{\pi}{2 K_2}$, $T_j=t_j$, $\rho=0$, $j=1,2$.

\section{Acknowledgments}

This research was supported by the Research Project of National Interest (PRIN) No. 2020X4T57A. It was also done within the activities of the INDAM-GNFM.


\begin{thebibliography}{99}

\bibitem{AL1} M. J. Ablowitz and J. F. Ladik, ``Nonlinear differential-difference equations'', \textit{Jour. Math. Phys.}, \textbf{16} (1975), pp. 598-603.
	
\bibitem{AL2} M. J. Ablowitz and J. F. Ladik, ``Nonlinear differential-difference equations and Fourier Analysis'', \textit{Jour. Math. Phys}., \textbf{17} (1976), pp. 1011-1018.

\bibitem{AM1} M.J. Ablowitz, Z.H. Musslimani, ``Integrable nonlocal nonlinear Schr\"odinger equation'', \textit{Phys. Rev. Lett.}, \textbf{110}:6 (2013), 064105; doi:10.1103/PhysRevLett.110.064105.

\bibitem{APT}  Ablowitz, M., Prinari, B., and Trubatch, A. (2003). ``Discrete and Continuous Nonlinear Schr\"odinger Systems (London Mathematical Society Lecture Note Series)''. Cambridge: Cambridge University Press. doi:10.1017/CBO9780511546709.

\bibitem{AS}  M.J.Ablowitz and H. Segur, \textit{Solitons and the Inverse Scattering Transform}, SIAM, Philadelphia, 1981.

\bibitem{Akhm_AL} N. Akhmediev and A. Ankiewicz, ``Modulation instability, Fermi-Pasta-Ulam recurrence, rogue waves, nonlinear phase shift, and exact solutions of the Ablowitz-Ladik equation'', \textit{Phys. Rev. E} {\bf 83} (2011), 046603 .

\bibitem{Akhmed0} N.N. Akhmediev, V.M. Eleonskii, and N.E. Kulagin, ``Generation of periodic trains of picosecond pulses in an optical fiber: exact solutions'', \textit{Sov. Phys. JETP}, \textbf{62}:5 (1985), 894--899.

\bibitem{BF} T.B. Benjamin, J.E. Feir, ``The disintegration of wave trains on deep water. Part I. Theory'', \textit{Journal of Fluid Mechanics}, \textbf{27}:3 (1967) 417--430; doi:10.1017/S002211206700045X.

\bibitem{Talanov} V. I. Bespalov, V. I. Talanov, ``Filamentary structure of light beams in nonlinear liquids'', \textit{JETP Letters}, \textbf{3}:12 (1966), 307-310.

\bibitem{CGS} F. Coppini, P. G. Grinevich and P. M. Santini: ``The effect of a small loss or gain in the periodic NLS anomalous wave dynamics. I'',  Phys. Rev. E {\bf 101} (2020), 032204 . DOI: 10.1103/PhysRevE.101.032204.

\bibitem{CS1} F. Coppini and P. M. Santini: ``The Fermi-Pasta-Ulam-Tsingou recurrence of periodic anomalous waves in the complex Ginzburg-Landau and in the Lugiato-Lefever equations'', \textit{Phys. Rev. E } {\bf 102}, 062207 (2020). DOI: 10.1103/PhysRevE.102.062207.

\bibitem{CGS2} Coppini F., Grinevich P.G., Santini P.M. (2022) Periodic Rogue Waves and Perturbation Theory. In: Meyers R.A. (eds) Encyclopedia of Complexity and Systems Science. Springer, Berlin, Heidelberg. https://doi.org/10.1007/978-3-642-27737-5$\_$762-1

\bibitem{CS2} F. Coppini and P. M. Santini: ``The Massive Thirring Model: exact solutions and Fermi-Pasta-Ulam-Tsingou recurrence of anomalous waves''. Preprint 2023 (in preparation).

\bibitem{CS3} F. Coppini and P. M. Santini: ``The effect of loss/gain and hamiltonian perturbations of the Ablowitz - Ladik lattice on the recurrence of periodic anomalous waves''. Preprint 2023  (in preparation). 

\bibitem{Doliwa}  A. Doliwa and P. M. Santini: ``Integrable dynamics of a discrete curve and the Ablowitz-Ladik hierarchy''; \textit{J. Math. Phys. }{\bf 36} (1995), 1259-1273.

\bibitem{Dubrovin} B.A. Dubrovin, ``Inverse problem for periodic finite-zoned potentials in the theory of scattering'', \textit{Funct. Anal. Appl.}, \textbf{9}:1 (1975), 61--62; doi:10.1007/BF01078183.

\bibitem{Dysthe} K.B. Dysthe, K. Trulsen, ``Note on Breather Type Solutions of the NLS as Models for Freak-Waves'', \textit{Physica Scripta}, \textbf{T82}  (1999) 48--52; doi:10.1238/Physica.Topical.082a00048.

\bibitem{FPU} G. Gallavotti (Ed.), ``The Fermi-Pasta-Ulam Problem: A Status Report'', \textit{Lecture Notes in Physics}, Vol. 728, Springer, Berlin Heidelberg, 2008; doi:10.1007/978-3-540-72995-2.

\bibitem{Geng}  Xianguo Geng, ``Darboux Transformation of The Discrete Ablowitz–Ladik Eigenvalue Problem'', \textit{Acta Mathematica Scientia}, \textbf{9} (1989), 1, 21-26.  doi:10.1016/S0252-9602(18)30326-6.


\bibitem{GS1} P.G. Grinevich, P.M. Santini, ``The finite gap method and the analytic description of the exact rogue wave recurrence in the periodic NLS Cauchy problem. 1'', \textit{Nonlinearity}, \textbf{31}:11 (2018), 5258--5308; doi:10.1088/1361-6544/aaddcf.

\bibitem{GS2} P.G. Grinevich, P.M. Santini: ``The finite-gap method and the periodic NLS Cauchy problem of anomalous waves for a finite number of unstable modes'', \textit{ Russian Math. Surveys} 74:2 211-263 (2019). DOI: https://doi.org/10.1070/RM9863.

\bibitem{GS3} P.G. Grinevich, P.M. Santini, ``The exact rogue wave recurrence in the NLS periodic setting via matched asymptotic expansions, for 1 and 2 unstable modes'', \textit{Physics Letters A}, \textbf{382}:14 (2018), 973--979; doi:10.1016/j.physleta.2018.02.014.

\bibitem{GS4} P.G. Grinevich, P.M. Santini: P. G. Grinevich and P. M. Santini, ``Numerical Instability of the Akhmediev Breather and a Finite-Gap Model of It'', In: V. Buchstaber, S. Konstantinou-Rizos, A. Mikhailov (eds), Recent Developments in Integrable Systems and Related Topics of Mathematical Physics. MP 2016. Springer Proceedings in Mathematics and Statistics, vol 273. Springer, Cham (Springer, Cham, 2018).

\bibitem{GS5} P.G. Grinevich, P.M. Santini: ``The linear and nonlinear instability of the Akhmediev breather'', Nonlinearity \textbf{34} (2021) 8331-8358. doi.org/10.1088/1361-6544/ac3143.
 
\bibitem{GS6} P.G. Grinevich, P.M. Santini: ``Phase resonances of the NLS rogue wave recurrence in the quasi-symmetric case'', \textit{Theoretical and Mathematical Physics}, \textbf{196}:3 (2018), 1294--1306; doi:10.1134/S0040577918090040.
  
\bibitem{HendersonPeregrine} K.L. Henderson, D.H. Peregrine, J.W. Dold, ``Unsteady water wave modulations: fully nonlinear solutions and comparison with the nonlinear Schr\"odinger equtation'', \textit{Wave Motion}, \textbf{29}:4 (1999), 341--361; doi:10.1016/S0165-2125(98)00045-6.

\bibitem{Hirota0} R. Hirota, Phys. Rev. Lett. {\bf 27}, 1192 (1971).

\bibitem{KharifPeli1} C. Kharif, E. Pelinovsky, ``Physical mechanisms of the rogue wave phenomenon'', \textit{Eur. J. Mech. B/ Fluids J. Mech.}, \textbf{22}:6 (2004), 603--634; doi:10.1016/j.euromechflu.2003.09.002.

\bibitem{KharifPeli2} C. Kharif, E. Pelinovsky, ``Focusing of nonlinear wave groups in deep water'' \textit{JETP Lett.}, \textbf{73} (2011), 170--175.

\bibitem{Kimmoun} O. Kimmoun, H.C. Hsu, H. Branger, M.S. Li, Y.Y. Chen, C. Kharif, M. Onorato, E.J.R. Kelleher, B. Kibler, N. Akhmediev, A. Chabchoub, ``Modulation Instability and Phase-Shifted Fermi-Pasta-Ulam Recurrence'', \textit{Scientific Reports}, \textbf{6} (2016), 28516. doi:10.1038/srep28516.

\bibitem{Krichever} I.M. Krichever, ``Methods of algebraic Geometry in the theory on nonlinear equations'', \textit{Russian Math. Surv.}, \textbf{32} (1977), 185--213; doi:10.1070/RM1977v032n06ABEH003862.

\bibitem{Kri} I.M. Krichever, ``The periodic non-Abelian Toda chain and its two-dimensional generalization''; Appendix of \cite{Dubrovin}.

\bibitem{Kuznetsov}  E. A. Kuznetsov, ``Solitons in a parametrically unstable plasma'', \textit{Sov. Phys. Dokl.}, \textbf{22} (1977), 507–508.

\bibitem{Infeld} E. Infeld, ``Quantitive theory of the {F}ermi-{P}asta-{U}lam recurrence in the {N}onlinear {S}chr{\"o}dinger {E}quation''. \textit{Phys. Rev. Lett.} \textbf{47} (1981), 717-718,

\bibitem{ishimori}
Y. Ishimori, ``An integrable classical spin chain'', \textit{J. Phys. Soc. Japan} \textbf{51} (1982), 3417–3418. 

\bibitem{ItsKorepin} A. R. Its, A. G. Izergin, V. E. Korepin, and N. A. Slavnov, ``Temperature correlations of quantum spins'', \textit{Phys. Rev. Letters}, \textbf{70}:11 (1991), 1704 - 1706.
  
 
\bibitem{ItsKotlj} A.R. Its, V.P. Kotljarov, ``Explicit formulas for solutions of a nonlinear Schr\"odinger equation'', \textit{Dokl. Akad. Nauk Ukrain. SSR Ser. A}, \textbf{1051} (1976), 965--968 .
 
\bibitem{ItsMatveev} A.R. Its, V.B. Matveev, ``Hill's operator with finitely many gaps'', \textit{Funct. Anal. Appl.}, \textbf{9}:1 (1975), 65--66; doi:10.1007/BF01078185.


\bibitem{ItsRybinSall} A.R. Its, A.V. Rybin, M.A. Sall, ``Exact integration of nonlinear Schr\"odinger equation'', \textit{Theor. Math. Phys.}, \textbf{74} (1988), 20--32; doi:10.1007/BF01018207.

\bibitem{Lieb} E. Lieb, T. Schultz, and D. Mattis, ``Two soluble models of an antiferromagnetic chain'', \textit{Ann. Phys. (N.Y.)} \textbf{16} (1961), 407--446 .

\bibitem{LL} L. A. Lugiato and R. Lefevre, ``Spatial Dissipative Structures in Passive Optical Systems'', \textit{Phys. Rev. Letters} \textbf{85} (1987), 2209--2211.

\bibitem{Ma}  Y. C. Ma, ``The perturbed plane wave solutions of the cubic Schr\"odinger equation'', Stud. Appl. Math., \textbf{60}:1 (1979), 43–58;
 
\bibitem{Marquie} P. Marqui\'e,  J. M. Bilbault,  M.  Remoissenet, ``Observation of nonlinear localized modes in an electrical lattice''. \textit{Physical Review E}, \textbf{51}:6 (1995), 6127–6133; doi:10.1103/physreve.51.6127.

\bibitem{Mikhailov} A. V. Mikhailov, ``Integrability of the two-dimensional Thirring model'', JETP Lett. (USSR) (Engl. Transl.) \textbf{23}:6 (1976).

\bibitem{Miller} P. D. Miller,  N. M. Ercolani,  I. M. Krichever and  C. D. Levermore, ``Finite genus solutions to the {A}blowitz--Ladik equations'', \textit{Communications on Pure and Applied Mathematics }, \textbf{48} (1995),  1369-1440.
	
\bibitem{Mussot} A. Mussot, C. Naveau, M. Conforti, A. Kudlinski, P. Szriftgiser, F. Copie, S. Trillo, ``Fibre multiwave-mixing combs reveal the broken symmetry of Fermi-Pasta-Ulam recurrence'', \textit{Nature Photonics}, \textbf{12}:5 (2018), 303--308; doi:10.1038/s41566-018-0136-1.

\bibitem{Narita} K. Narita, ``Soliton Solutions for Discrete Hirota Equation. II'', J. Phys. Soc. Jpn., {\bf 60}:5 (1991), 1497–1500; doi:10.1143/JPSJ.60.1497.

\bibitem{Newell_Whitehead} A. C. Newell and J. A. Whitehead,  ``Review of the Finite Bandwidth Concept'',  Proc. I.U.T.A.M.\textit{ Symposium on Instability of Continuous Systems}, Springer-Verlag, Berlin, 1969, pp. 284-289; doi:10.1007/978-3-642-65073-4\_39.  

\bibitem{Novikov} S.P. Novikov, ``The periodic problem for the Korteweg-de Vries equation'', \textit{Funct. Anal. Appl.}, \textbf{8}:3 (1974), 236--246; doi:10.1007/BF01075697.


\bibitem{Onorato2} M. Onorato, S. Residori, U. Bortolozzo, A. Montina, F.T. Arecchi, ``Rogue waves and their generating mechanisms in different physical contexts'', \textit{Physics Reports}, \textbf{528}:2 (2013), 47--89; doi:10.1016/j.physrep.2013.03.001.

\bibitem{Osborne} A. Osborne, M. Onorato, M. Serio, ``The nonlinear dynamics of rogue waves and holes in deep-water gravity wave trains'', \textit{Phys. Lett. A}, \textbf{275}(5-6) (2000), 386--393; doi:10.1016/S0375-9601(00)00575-2.

\bibitem{Otha} Y. Ohta and J. Yang, ``General rogue waves in the focusing and defocusing Ablowitz-Ladik equations'', \textit{J. Phys. A: Math. Theor.} \textbf{47} (2014), 255201 (23pp).

\bibitem{Peregrine} D.H. Peregrine, ``Water waves, nonlinear Schr\"odinger equations and their solutions'', \textit{J. Austral. Math. Soc. Ser. B}, \textbf{25} (1983), 16--43; doi:10.1017/S0334270000003891

\bibitem{PieranFZMAGSCDR} D. Pierangeli, M. Flammini, L. Zhang, G. Marcucci, A.J. Agranat, P.G. Grinevich, P.M. Santini, C. Conti, E. DelRe, ``Observation of exact Fermi-Pasta-Ulam-Tsingou recurrence and its exact dynamics'', \textit{Phys. Rev. X}, \textbf{8}:4 (2019), p. 041017 (9 pages); doi:10.1103/PhysRevX.8.041017.

\bibitem{Prinari} B. Prinari, ``Discrete solitons of the focusing Ablowitz-Ladik equation with nonzero boundary conditions via inverse scattering''. \textit{Journal of Mathematical Physics}, \textbf{57}(8) (2016), 083510. doi:10.1063/1.4961160 

\bibitem{San} P.M. Santini, ``The periodic Cauchy problem for PT-symmetric NLS, I: the first appearance of rogue waves, regular behavior or blow up at finite times'', \textit{J. Phys. A: Math. Theor.}, \textbf{51}:49 (2018), 495207 (21pp); doi:10.1088/1751-8121/aaea05.

\bibitem{RungeKutta} D. Sarafyan, ``Improved sixth-order Runge-Kutta formulas and approximate continuous solution of ordinary differential equations'', \textit{Journal of Mathematical Analysis and Applications}, {\bf 40} (1972), 436-445. doi:10.1016/0022-247X(72)90062-5.

\bibitem{Schober}
C. M. Schober, ``Numerical and analytical studies of the discrete nonlinear Schroedinger equation'', PhD  Thesis, University of Arizona (1991).

\bibitem{Solli} D.R. Solli, C. Ropers, P. Koonath and B. Jalali, ``Optical rogue waves'', \textit{Nature}, \textbf{450} (2007), 1054--1057; doi:10.1038/nature06402.

\bibitem{Soto} J.M. Soto-Crespo, N. Devine, and N. Akhmediev, ``Adiabatic transformation of continuous waves into trains of pulses'', \textit{Phys. Rev. A} \textbf{96}, 023825 (2017).

  
\bibitem{Stokes} G. Stokes, ``On the Theory of Oscillatory Waves'', \textit{Transactions of the Cambridge Philosophical Society} \textbf{VIII} (1847), 197--229, and Supplement 314--326.
\bibitem{Takeno}
S. Takeno and K. Hori, ``A propagating self-localized mode in a one-dimensional lattice with quartic anharmonicity'', \textit{J. Phys. Soc. Japan}, \textbf{59} (1990), 3037–3040 .

\bibitem{Thirring} W. E. Thirring, ``A soluble relativistic field theory''. \textit{Annals of Physics}, \textbf{3} (1958), 91–112. doi:10.1016/0003-4916(58)90015-0

  
\bibitem{trillo3} S.~Trillo and S.~Wabnitz, ``Dynamics of the nonlinear modulational instability in optical fibers'', \textit{Optics Letters}, \textbf{16} (13): 986-988, 1991.

\bibitem{Simaeys} G. Van Simaeys, P. Emplit, M. Haelterman, ``Experimental Demonstration of the Fermi-Pasta-Ulam Recurrence in a Modulationally Unstable Optical Wave'', \textit{Phys. Rev. Lett.}, \textbf{87}:3 (2001), 033902; doi:10.1103/PhysRevLett.87.033902.

\bibitem{Yuen1} H.C. Yuen, W.E. Ferguson, ``Relationship between Benjamin-Feir instability and recurrence in the nonlinear Schr\"odinger equation'', \textit{Phys. Fluids}, \textbf{21}:8 (1978), 1275--1278; doi:10.1063/1.862394.

\bibitem{Yuen3} H. Yuen, B. Lake, ``Nonlinear dynamics of deep-water gravity waves'', \textit{Advances in Applied Mechanics}, \textbf{22} (1982), 67--229; doi:10.1016/S0065-2156(08)70066-8.


\bibitem{Yurov} A.V. Yurov, and V.A. Yurov, ``The Landau-Lifshitz Equation, the NLS, and the Magnetic Rogue Wave as a By-Product of Two Colliding Regular ``Positons"'', Symmetry, {\bf 10} (2018), 82. doi:10.3390/sym10040082.


\bibitem{Zakharov} V.E. Zakharov, ``Stability of period waves of finite amplitude on surface of a deep fluid'', \textit{Journal of Applied Mechanics and Technical Physics}, \textbf{9}:2 (1968) 190--194.

\bibitem{ZMNP}  V.E. Zakharov, S.V. Manakov, S.P. Noviko and L.P. Pitaevsky, 1984 \textit{Theory of Solitons}, (New York: Plenum).

\bibitem{ZakharovOstro} V. Zakharov, L. Ostrovsky, ``Modulation instability: the beginning'', \textit{Physica D: Nonlinear Phenomena}, \textbf{238}:5 (2009), 540--548; doi:10.1016/j.physd.2008.12.002.


\bibitem{ZakharovShabat} V.E. Zakharov, A.B. Shabat, ``Exact theory of two-dimensional self-focusing and one-dimensional self-modulation of waves in nonlinear media'', \textit{Sov. Phys. JETP}, \textbf{34}:1 (1972), 62--69.

\end{thebibliography}
\end{document}